\documentclass[letterpaper,12pt]{article}
\usepackage{graphicx}
\usepackage{hyperref}
\usepackage{amsthm,amsmath,amsfonts,amssymb}
\usepackage{color}
\usepackage[authoryear,comma,longnamesfirst,sectionbib,square]{natbib}
\usepackage[normalem]{ulem}
\usepackage{subcaption}

\newcommand{\beq}{\begin{equation}}
\newcommand{\eeq}{\end{equation}}

\newcommand{\cI}{\mathcal I}

\newcommand{\cP}{\mathcal P}
\newcommand{\cC}{\mathcal C}
\newcommand{\cS}{\mathcal S}

\newcommand{\SoS}{\mbox{SoS}}
\newcommand{\SoP}{\mbox{SoP}}

\newcommand{\CoS}{\mbox{CoS}}

\newcommand{\FCR}{\mbox{FCR}}

\newcommand{\rank}[1]{{(#1)}}

\newcommand{\Sidak}{\textnormal{\v{S}id}}

\newcommand{\ubar}[1]{\text{\b{$#1$}}}

\def\EE{\mathbb{E}}

\renewcommand{\Pr}{\mathbb{P}}

\newtheorem{theorem}{Theorem}
\newtheorem{prop}[theorem]{Proposition}
\newtheorem{corollary}[theorem]{Corollary}

\newtheorem{remark}[theorem]{Remark}

\newtheorem{rem*}{Remark}

\begin{document}

\title{Confidence Intervals for Selected Parameters}

\author{Yoav Benjamini\footnote{%
	Authors are listed alphabetically.} 
	\footnote{%
    The research leading to these results has received funding from the European Research Council under the European Community’s Seventh Framework Programme (FP7/2007-2013) / ERC grant agreement n [294519] PSARPS.} \\
        Department of Statistics and Operations Research \\
        The Sackler School of Mathematical Sciences \\ and \\
        Sagol School of Neurosciences\\
        Tel Aviv University\\[.1in]
        Yotam Hechtlinger\\ 
        Department of Statistics and Data Science \\
        Carnegie Mellon University\\[.1in]
	Philip B.~Stark \\
	Department of Statistics\\
	University of California, Berkeley
}

\date{}

\maketitle

\newpage

\begin{abstract}
    Practical or scientific considerations often lead to selecting a subset of parameters
    as ``important.''
    Inferences about those parameters often are based on the same data
    used to select them in the first place.
    That can make the reported uncertainties deceptively optimistic:
    confidence intervals that ignore selection generally have less than their nominal
    coverage probability.
    Controlling the probability that one or more intervals for selected parameters 
    do not cover---the ``simultaneous over the selected'' (\SoS{}) error rate---is crucial in many scientific problems.
    Intervals that control the \SoS{} error rate can be constructed in ways that take
    advantage of knowledge of the selection rule.
    We construct \SoS{}-controlling confidence
    intervals for parameters deemed the most ``important''
    $k$ of  $m$ shift parameters because they are estimated (by independent estimators)
    to be the largest.
    The new intervals improve substantially over \v{S}id\'{a}k intervals when $k$ is small compared to $m$,
    and approach the standard Bonferroni-corrected intervals when $k \approx m$.    
    Standard, unadjusted confidence intervals for location parameters have the correct coverage 
    probability for $k=1$, $m=2$
    if, when the true parameters are zero, the estimators are exchangeable and symmetric.
\end{abstract}

\noindent%
\vfill
\section{Introduction} \label{sec:introduction}
Modern statistical applications, for instance, studies of high-throughput experiments or 
high-dimensional databases, rarely involve only one parameter. 
Rather, investigators generally rely on model selection or parameter selection of 
some kind, e.g., algorithmically, by trial and error, or merely by examining 
the data before deciding what analysis to perform. 
Practical or scientific reasons may limit interest or reporting to a subset of parameters, even 
when the analysis involved many more.
These may be the only parameters reported in the analysis, or simply emphasized above the rest
by inclusion in the abstract, discussion, tables, or figures. 
Inferences about such selected parameters often are based on the same data used to decide they are ``important.''

Because selection alters the distribution of $p$-values, estimators, and test statistics,
selection complicates inference.
The ASA board  issued a warning about $p$-values for selected 
parameters~\citep{wasserstein2016ASA}, but did not suggest any remedy.
Instead, they recommended using alternatives such as confidence intervals.
Unfortunately, selection also alters the coverage probability of standard confidence intervals.
Here, we provide new methods for constructing simultaneous confidence intervals 
for parameters selected because their estimates are largest.
The methods illustrate that using information about the specific way the parameters
are selected can improve inferences.

The issue of inference after selecting hypotheses, parameters, or 
models for inference after observing the data---\emph{selective inference}---was recognized 
70 years ago in the problem of making all pairwise comparisons between independent groups.
The field that copes with selective inference came to be known
as `multiple comparisons.'
Inference about the difference selected as interesting because it was estimated to be largest 
led Tukey to introduce the studentized range \citep{tukey1953theProblem,braun1994collected},
and selection has remained a motivating theme, as exemplified by the introduction to 
the definitive work by \citet{hochberg1987multiple}.
That motivation notwithstanding, the book's solutions involved simultaneous coverage over 
\emph{all} parameters of potential interest, thereby controlling the error probability for 
whatever subset was selected. 
Similarly, \citet{berk2013valid}, who develop 
methods for selective inference in the context of model selection in linear regression, rely on 
simultaneous protection.

Addressing selective inference by simultaneous coverage
of all possible subsets of parameters is conservative in ``modern'' statistical
problems with very many potential inferences, often more than 
observations (``$p \gg n$''). 
\citet{benjaminiYekutieli05} distinguished between 
\emph{simultaneous inference} and \emph{selective inference}.
The latter requires that the inferential property (e.g., the Type~I error rate or coverage probability) 
hold \emph{on average over the selected parameters}.
Selective inference is often less stringent than simultaneous inference.
For instance, if a collection of intervals has simultaneous coverage probability $1-\alpha$, 
its selective coverage is at least $1-\alpha$, but the converse is not true. 
Methods for selective inference can be more powerful than methods for simultaneous inference,
as illustrated by methods that control the False Coverage Rate (FCR) 
\citep{benjaminiYekutieli05}.

FCR control is a property of the entire selected set of intervals.
FCR is not controlled for subsets of \emph{that} set. 
This is not a hypothetical problem.
For instance, we might evaluate a large number of potential drug molecules for efficacy,
then decide to look more closely at the most promising ten.
One or two of the ten (not necessarily the largest) are then used in additional experiments. 
Regions on the genome may be identified to be associated with some disease, 
but only one or two locations in each region used for replication.
Identifying a few peaks of activity in fMRI studies is another example where
initial screening is followed by study of a subset of items that pass the initial screening. 
In all these examples, controlling FCR in the initial screening 
guarantees nothing about subsets of the set that pass.

This problem is attracting attention, but proposed solutions require specifying
how the sub-selection is performed.
See, e.g., \cite{katsevich2018towards} and references therein.
On the other hand, starting with simultaneous confidence intervals
could be too conservative or exaggerate the potential of treatments by including larger
effects than the data support.

In this work, we explore a middle way: 
Given a specific selection rule, a set of confidence intervals has 
\emph{simultaneous coverage over the selected} if the probability of not covering 
any selected parameter is controlled at a desired level.
Simultaneity is guaranteed only on the selected set, not all possible sets.
This allows us to make sharper inferences than omnibus protection against all selection rules allows.
For example, an interval for the parameter estimated to be the larger of two
does not need to be longer than the standard univariate interval; see Section~\ref{sec:larger_of_2}.
 
We demonstrate the advantage of this approach for a simple yet 
practical rule: ``select the $k$ parameters estimated to be largest.'' 
This rule is used---sometimes silently---in applications in which
a single risk factor is of primary interest, but there is a collection of possible confounders.
We also give a new result for the rule ``select the parameter estimated to be largest in 
absolute value'' in the bivariate normal case.


\section{Simultaneous over selected parameters}

We observe
$Y = (Y_i)_{i=1}^m$, where $Y_i \sim F_{\theta_i}^i$, $i = 1, \ldots, m$, are
real-valued random variables.
A \emph{selection rule} $S(\cdot)$ is a mapping from
$\Re^m$ into $\cP(\{1, \ldots, m\}) \equiv \cP(m)$,
the power set of $\{1, \ldots, m\}$,
that is $F_\theta$-measurable for all $\theta$.
The set $S(Y)$ consists of the indices of the \emph{selected parameters}:
If $Y=y$, we seek finite-length confidence intervals for 
$\{ \theta_i \}_{i \in S(y)}$. 
We make no confidence statement about the other
parameters.

The \emph{Simultaneous over all Possible selections} (\SoP{}) error rate is
\beq
   \SoP{} \equiv \max_{s \in \cP(m)} \EE \max_{i \in s} 1_{\theta_i \notin \cI_i(Y)},
\eeq
The \emph{False Coverage-statement Rate} (\FCR{}) 
controls non-coverage on average over the selected parameters:
\beq
     \FCR{} \equiv \EE_\theta \left [
     \frac{\sum_{i \in S(Y)} 1_{\theta_i \notin \cI_i(Y)}}{|S(Y)|} \right ],
\eeq
where $\frac{0}{0} \equiv 0$ (if no interval is constructed, no interval fails to cover).
Another error rate that explicitly involves $S$ is the \emph{Conditional over Selected} (\CoS{}):
\beq
    \CoS{} \equiv \EE_\theta \left [ \max_{i \in S(Y)} 1_{\theta_i \notin \cI_i(Y)} | i \in S(Y) \right ].
\eeq
(For a recent review of this criterion, see \cite{taylor2015statistical}.)

This paper focuses on the \emph{Simultaneous over Selected parameters} (\SoS{}) error rate:
\beq\label{eq:SoS_definition}
     \SoS{} \equiv \Pr \left \{  \exists  i \in S(Y) : \cI_i(Y) \not\ni \theta_i  \right \}.
\eeq
\SoS{} is the probability that any interval for a selected 
parameter fails to cover.
Controlling \SoS{} allows further selection from $S(y)$ without 
requiring the intervals to be modified.

Despite how often practitioners use the same data to select a set of parameters and then make 
inferences about the selected parameters, there are few results regarding controlling \SoS{}.
\citet{venter1988confidence} constructed a confidence interval for the mean estimated 
to be the largest among $m$ independent Gaussian estimators and
\citet{fuentes2018confidence} recently constructed \SoS{} intervals
for the $k$ means estimated to be largest among $m$ independent normal estimators.
``Multiple comparisons with sample best'' 
\citep{hsu81,hsu96} also involves testing a set of hypotheses 
that depends on the data through 
the identity of the parameter estimated to be largest,
to which all other parameter estimates are compared: 
this might be the first example of concern about \SoS{}. 
But the general idea of simultaneous coverage over a selected subset does not
seem to have been recognized, aside from the work of
\citet{HechtlingerYotam2014CIft}, incorporated here.

Controlling \SoP{} obviously controls both \SoS{} and \FCR{}. 
Controlling \CoS{} assures conditional coverage for each selected parameter, so it 
controls coverage on average for the selected parameters.
Therefore, \CoS{} controls the conditional \FCR{}, given that at least one parameter was selected.
Since conditional \FCR{} is larger than \FCR{}, controlling \CoS{} controls \FCR{}.

For some selection rules, intervals that control \CoS{} might not control \SoP{}; 
for others, \CoS{} yields longer intervals than \SoP{}. 
Intervals with $\CoS \le 1-\alpha/m$ control \SoP{} at level $1-\alpha$.

When $|S(Y)|=1$ with probability $1$, \SoS{} and \CoS{} are equivalent and \CoS{} intervals control \SoS{} and \SoS{} intervals control \CoS{}. 
If  $|S(Y)|\ge1$ with probability $1$,  $1-\alpha/|S(Y)|$ \CoS{} intervals control \SoS{} at level $1-\alpha$.

\section{Controlling \SoS{} by inverting non-equivariant unconditional tests} \label{sec:larger_of_2}

One way to construct intervals that control \SoS{} is to make simultaneous intervals for all
$m$ parameters. 
If we construct the simultaneous confidence intervals by inverting suitable non-equivariant hypothesis 
tests, then projecting the joint confidence set onto the selected components, we
can get ``selective'' behavior by designing the tests so that the confidence intervals
for the parameters that are \emph{not} selected are $(-\infty, \infty)$, in effect not making
any inference about those parameters.

For instance, consider the acceptance region $A_{\mu, c_\mu}$ 
for testing the hypothesis $\theta = \mu = (\mu_i)_{i=1}^m$:
\beq \label{eq:acceptance_regions_definition}
    A_{\mu, c_\mu} \equiv \{y \in \Re^m : |\mu_i - y_i | \le c_\mu; \;  i \in S(y)\},
\eeq
where 
\beq \label{eq:cm_mu_def}
c_\mu \equiv \inf_{c \in \Re}  \left \{c : \Pr_{\mu} \{ Y \notin A_{\mu, c_\mu}\} \le \alpha \right \}.
\eeq
Inverting this family of tests for $Y = y$ yields
\beq \label{eq:inversion_set}
	\cC(y)= \{ \mu \in \Re^m : y \in A_{\mu, c_\mu} \},
\eeq
which satisfies
\beq
 \Pr_{\theta} \{ \cC(Y) \not \ni \theta \} \le \alpha.
\eeq
Whenever $\cC(Y) \ni \theta$, $\theta_i$ is between $\inf_{\mu \in \cC(y)} \mu_i$
and $\sup_{\mu \in \cC(y)} \mu_i$, for all $i$.
Therefore, the intervals
\beq \label{eq:inversion_projection}
	\mathcal{I}_i (y) \equiv  \left [ \inf_{\mu \in \cC(y)} \mu_i, 
       \sup_{\mu \in \cC(y)} \mu_i \right ], \;\; i=1, \ldots, m,
\eeq
are simultaneous $1-\alpha$ confidence intervals for $\{\theta_i\}_{i=1}^m$.
Because these intervals have simultaneous $1-\alpha$ coverage for \emph{all} 
$\{\theta_i\}_{i=1}^m$, they have simultaneous coverage for $\{\theta_i\}_{i \in S(Y)}$.

This choice of $A_{\mu, c_\mu}$ gives these confidence intervals a selective structure:
Suppose $i \notin S(Y)$. 
Consider $\mu^\gamma$ with components
\beq
   \mu_i^\gamma \equiv  \begin{cases}
        Y_i, & i \in S(Y) \\
        \gamma, & i \notin S(Y) .
        \end{cases}
\eeq
Then $A_{\mu^\gamma, c_{\mu^\gamma}} \ni Y$.
Hence, $\mu^\gamma \in C(Y)$, and so $\gamma \in \cI_i(Y)$ for all $i \notin S(Y)$.
Since this holds for all $\gamma \in \Re$, $\mathcal{I}_i (Y) = (-\infty, \infty)$ for $i \notin S(Y)$:
the procedure does not try to constrain non-selected parameters.

In general, $c_\mu$ depends on the selection rule $S(\cdot)$, but bounds on $c_\mu$ 
can make it easy to invert these families of tests conservatively. 
For instance, if $\forall \mu$, $c_\mu \le c^+$, then
$A_{\mu,c_\mu} \subset A_{\mu, c^+}$, and 
$\mathcal{I}_i(y) \subseteq [y_i-c^+, y_i+c^+ ]$ for $i \in S(Y)$.
If one can find a simple function $f(\mu)$ such that $c_\mu \le f(\mu)$ for all $\mu$, then
conservative confidence intervals can be found by inverting acceptance regions based on $f$
rather than $c$.

\subsection{The larger of two exchangeable, symmetric estimates}\label{sec:max_2}
We construct \SoS{} intervals for the parameter $\theta_i$ estimated to be the larger of two:
$S(y) \equiv \{ \rank{1}\}$, where $\rank{1}$ denotes the index of the larger of $Y_1$ and $Y_2$.
(Ties can be broken lexicographically.)
We specialize to the case $\{Y_i-\theta_i\}_{i=1}^2$ are exchangeable and symmetric;
they need not be independent or continuous.
This generalizes a result for correlated bivariate normals with unit variance given by \citet{HechtlingerYotam2014CIft}.

Define the set transformation
\beq \label{eq:minusTdef}
   A^{-T} \equiv \{(-y_2, -y_1) : (y_1, y_2) \in A \}.
\eeq
Symmetry and exchangeability imply that if $A \subset \Re^2$ is measurable
with respect to the joint distribution of $\{Y_i\}$, so is $A^{-T}$,
and $\Pr_0 \{A^{-T}\} = \Pr_0 \{A\}$. 
By definition, $\Pr_\mu \{A\} = \Pr_0 \{A-\mu\}$.
  
For this $S$, the acceptance region \eqref{eq:acceptance_regions_definition} 
is the union of two disjoint semi-infinite trapezoids:
\beq
	A_{\mu,c}\equiv A_1 \cup A_2
\eeq
where
\begin{align*}
    A_1 & \equiv \{ y \in \Re^2 : | y_1 - \mu_1 | \le  c \mbox{ and }
                                           y_1 \ge  y_2 \} \\
    A_2 & \equiv \{ y \in \Re^2 : | y_2 - \mu_2 | \le  c \mbox{ and }
                                           y_2 >  y_1 \}.
\end{align*}

\begin{prop} \label{prop:exchangeable_symmetric}
Suppose $(Y_i - \theta_i)$, $i = 1, 2,$ are exchangeable with a symmetric marginal 
distribution, and define
$c_\alpha \equiv -\sup \{ c: \Pr \{ Y_i-\theta_i \le c \} \le \alpha \}$. 
For all $\mu \in \Re^2$,
   \[
      \Pr_\mu \{ Y \in A_{\mu, c_{\alpha/2}} \} \ge 1-\alpha.
    \]
\end{prop}

\begin{proof}
The proof hinges on the fact that
\beq \label{eq:unionIsSlab}
   A_1 \cup \left \{ (A_2 - \mu)^{-T} + \mu \right \} = 
   \{y \in \Re^2: |y_1 - \mu_1 | \le c \}.
\eeq
\emph{The LHS is a subset of the RHS:}
If $y \in  A_1$, $|y_1 - \mu_1 | \le c$.
If $y \in (A_2 - \mu)^{-T} + \mu$,
there exists $x$ with $x_2 > x_1$ and $|x_2 - \mu_2| \le c$ for which
$y = (x-\mu)^{-T} + \mu = ( -x_2+\mu_2 + \mu_1, -x_1 + \mu_1 + \mu_2)$.
Thus $ |y_1 - \mu_1| = |-x_2 + \mu_2| \le c$, and $y_1 - y_2 = x_1 - x_2 > 0$,
so $y \in A_1$.
\\[1ex]
\emph{The RHS is a subset of the LHS:}
If $|y_1 - \mu_1| \le c$ and $y_1 \ge y_2$, $y \in A_1$.
If $|y_1 - \mu_1| \le c$ and $y_1 < y_2$,  
$x = (y - \mu)^{-T} + \mu = (-y_2 + \mu_2 + \mu_1, -y_1 + \mu_1 + \mu_2)$.
Then $|x_2 - \mu_2| = | -y_1 + \mu_1| \le c$, and $x_2 - x_1 = -y_1 + y_2 > 0$, so
$x \in A_2$.
Calculation verifies $(x-\mu)^{-T}+\mu = y$. Thus \eqref{eq:unionIsSlab} holds.

Now 
\begin{align*}
  \Pr_\mu A_{\mu,c} & = \Pr_0 \{A_{\mu,c} - \mu \} \\
      & = \Pr_0 \{ (A_1 - \mu) \cup (A_2 - \mu) \} \\
      &=  \Pr_0 \{A_1 - \mu \} + \Pr_0 \{A_2 - \mu\} \\
      &= \Pr_0 \{A_1 - \mu \} + \Pr_0 \{(A_2 - \mu)^{-T}\} \\
      &= \Pr_0 \{ (A_1 - \mu)  \cup (A_2 - \mu)^{-T} \} \\
      & = \Pr_\mu \left \{ A_1  \cup \left \{ (A_2 - \mu)^{-T} + \mu \right \} \right \} \\
      & = \Pr_\mu \{ Y \in \{y: |y_1-\mu_1| \le c\} \}\\
      &= \Pr_0 \{ | Y_1  |  \le c \}.
\end{align*}
The conclusion follows by substituting $c = c_{\alpha/2}$.
\end{proof}

\begin{figure}[ht]
\centering
\includegraphics[width=0.30\paperwidth]{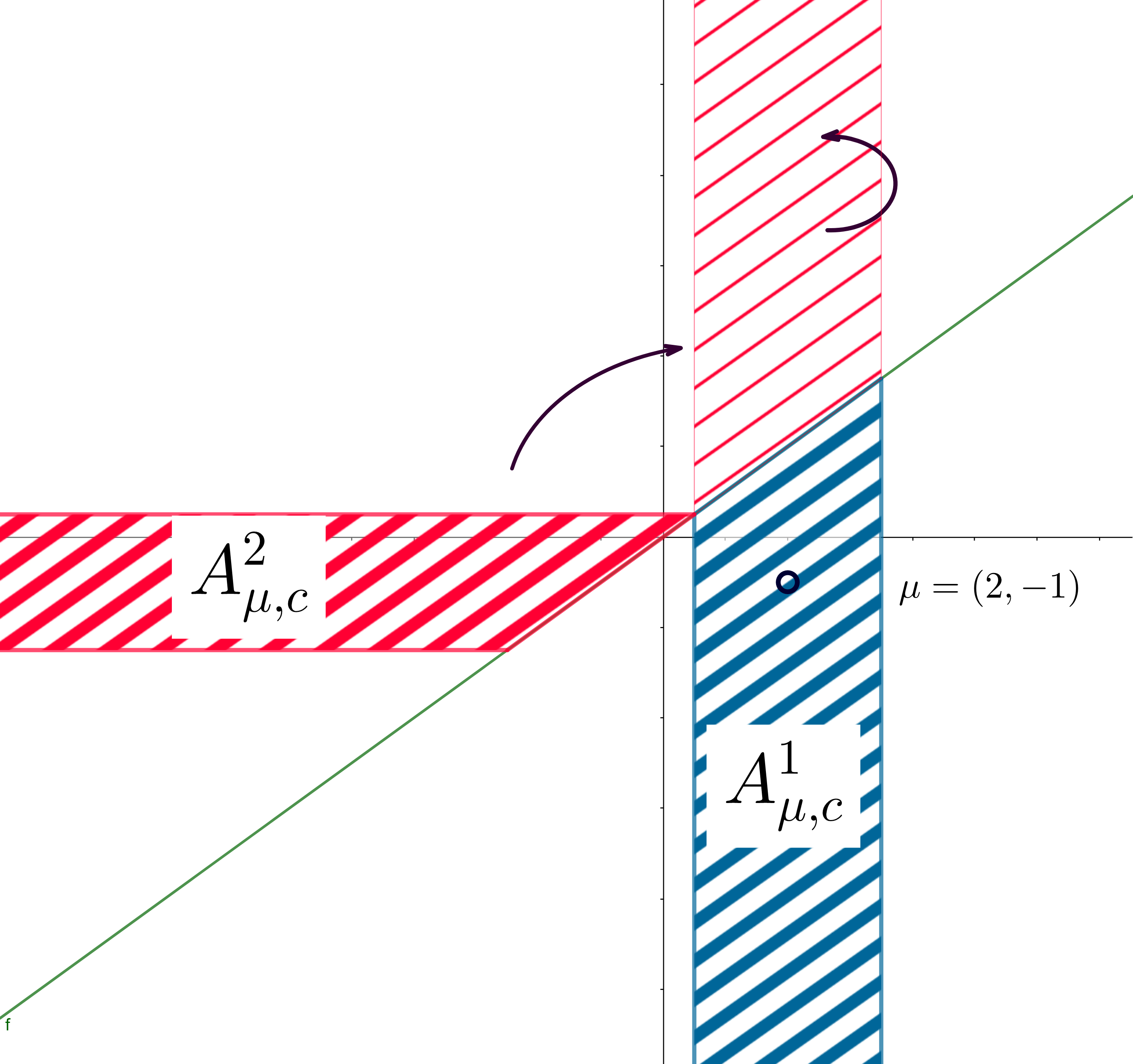}
\caption{ \protect \label{fig:2DTrapMaxTrans}
The semi-infinite trapezoids that comprise the acceptance region $A_{\mu,c}$. 
The blue hashed region is $A_{\mu_1,c}^1$ and the dark red hashed region is $A_{\mu_2,c}^2$. 
Exchangeability and symmetry of $\{Y_i - \theta_i\}$ imply that 
the probability of the lighter red region is equal to that of the dark red region if $\theta = \mu$,
for every $\mu \in \Re^2$.
}
\end{figure}

\noindent
Figure~\ref{fig:2DTrapMaxTrans}~(a) illustrates the key idea in the proof.

The proposition shows that $A_{\mu, c_{\alpha/2}}$ is the acceptance
region for a level $\alpha$ test of the hypothesis $\theta = \mu$.
As mentioned above, inverting tests of this form yields the confidence interval
\beq
  \cI(y) \equiv [y_\rank{1} -  c_{\alpha/2}, y_\rank{1} + c_{\alpha/2}].
\eeq
This is the standard two-sided univariate symmetric confidence interval, based on
the larger of the two observations. 
\emph{The standard two-sided univariate confidence
interval has the right coverage probability despite the selection and the dependence.}
\subsection{Larger absolute value of two normal estimators}\label{sec:abs_max_2}

This section addresses constructing a confidence interval for the component selected by the
rule $S(y) = \arg \max_i |y_i|$, where $Y \sim N(\theta, I)$
with $\theta \in \Re^2$.
(Since the two estimators have normal distributions, the probability of a tie is zero.)
As above, we find the intervals by inverting a family of hypothesis tests.
The acceptance regions for the tests are again a union of two pieces, but
they are more complicated than those in section~\ref{sec:max_2}.

\subsubsection{Acceptance regions}
\begin{figure}
\centering
\includegraphics[width=0.8\textwidth]{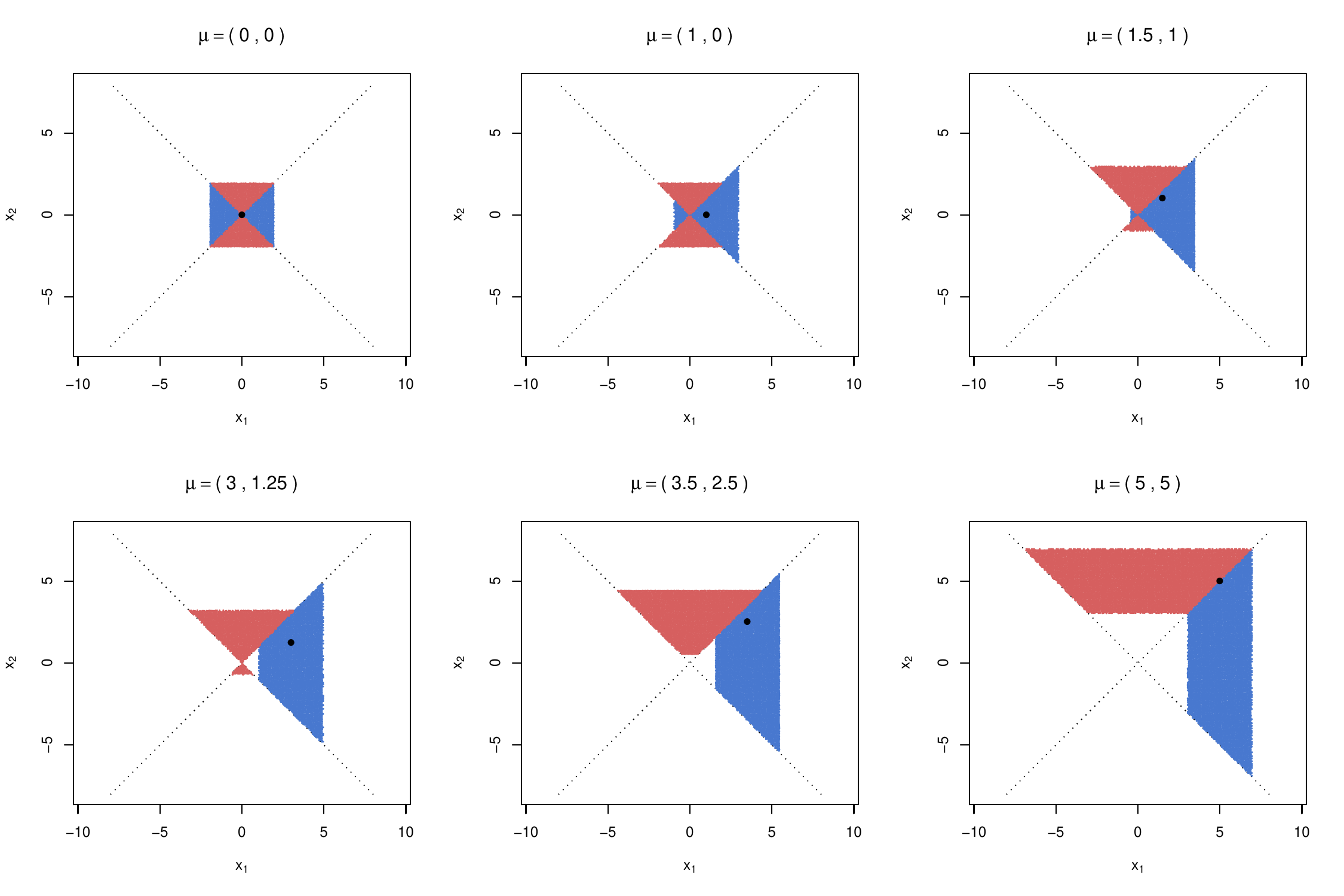}

\caption{ \protect \label{fig:max_abs_of_2_test}
$B_{\mu, c}$ for various values of $\mu$ (the black dot in the figures). The blue area is $B_{\mu_1, c}^1$ and the cayenne area is $B_{\mu_2,c}^2$.}
\end{figure}

The acceptance region $B_{\mu,c}$ for testing the hypothesis $\theta=\mu$ is
\beq
   B_{\mu, c_\mu} \equiv B_{\mu_1,c_\mu}^1 \cup B_{\mu_2,c_\mu}^2,
\eeq
where
\beq
  B_{\mu_i, c_\mu}^i \equiv 
     \left\{ 
         y \in \Re^2 : |\mu_i - y_i | \le c_\mu \mbox{ and } |y_i| > |y_j|, j \ne i
     \right\},
\eeq
and $c_\mu$ is chosen so that the test has level $\alpha$.
Figure~\ref{fig:max_abs_of_2_test} plots acceptance regions for 
six values of $\mu$.

\subsubsection{Finding $c_\mu$}

The constant $c_\mu$ is the smallest $c$ for which 
$\Pr_\mu \left \{ Y \in B_{\mu, c} \right \} \ge 1-\alpha$.
At $\mu = 0$, $B_{0, c_0}$ is a square;
so, $c_0$ is \v{S}id\'{a}k's constant $c_\Sidak$.
We will bound $c_\mu$ for $\mu \ne 0$.

\begin{prop} \label{prop:max_abs_of_2_bound}
Let $a \equiv \max(|\mu_1|,|\mu_2|)$.
For for all $c>0$,
\beq
     \Pr_\mu \left \{ Y \in  B_{\mu, c} \right \}
     \ge \Pr_{(a,0)} \left \{ Y \in  B_{(a, 0), c} \right \}.
\eeq
\end{prop}
\noindent
The proof is in appendix \ref{app:max_abs_of_2_proof}. Define 
\beq
   c_\mu^+ \equiv c_{(|\mu_1| \vee |\mu_2|, 0)} = c_{(0, |\mu_1| \vee |\mu_2|)}.
\eeq
By Proposition~\ref{prop:max_abs_of_2_bound},
$c_\mu^+ \ge c_\mu$. Explicit calculation gives
\beq
 \Pr_\mu \left \{ Y \in B_{\mu, c_\mu^+} \right \} =
  \sum_{i \in \{ 1, 2 \} } 
     \intop_{-c}^c \phi(t) 
         \left[
              \Phi(t+\mu_i - \mu_{i'}) - \Phi(-t-\mu_i-\mu_{i'}) 
         \right](-1)^{1_{ t+ \mu_i <0 }}
        dt, \label{eq:c_plus_integral}
\eeq
where $i' = 3-i$.
The value of $c_\mu^+$ is the smallest $c$ for which \eqref{eq:c_plus_integral}
is at least $1-\alpha$.

\begin{figure}
\centering
\includegraphics[scale=0.4]{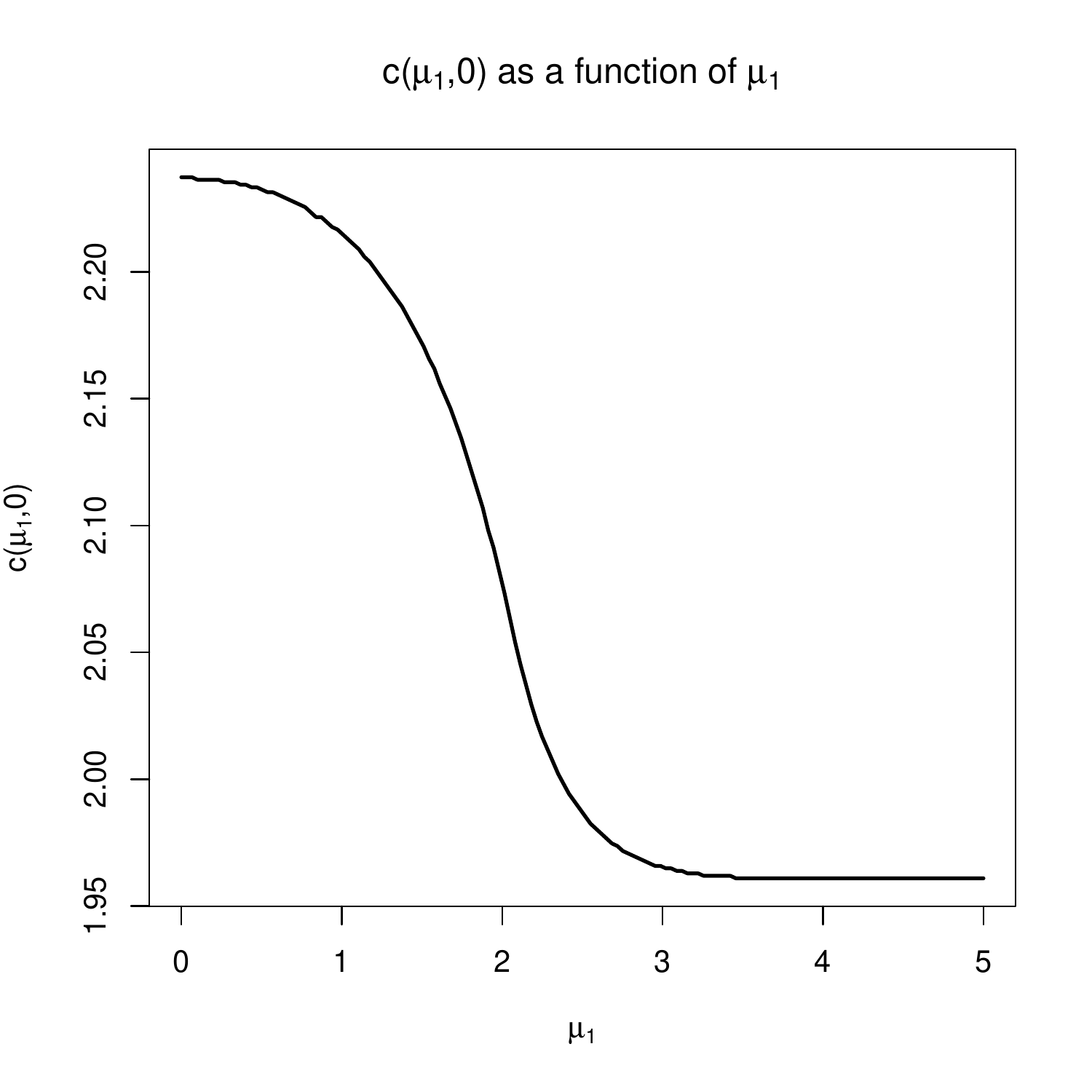}
\caption{\protect \label{fig:max_abs_of_2_bound} The upper bound $c_\mu^+$ as a function of $\max( |\mu_1|, |\mu_2|)$.
}
\end{figure}

Hence, tests using $c_\mu^+$ instead of $c_\mu$ have level no larger than $\alpha$, 
and inverting them will give confidence intervals with coverage probability at least $1-\alpha$.
Figure~\ref{fig:max_abs_of_2_bound} plots $c_\mu^+$ as a function of $\max (|\mu_1|, |\mu_2|)$.

As $\max( |\mu_1|, |\mu_2|)$ grows, $c_\mu^+$ decreases monotonically;
$c_{(3,0)} \approx z_{1-\frac{\alpha}{2}}$, the half-width
of a standard univariate normal confidence interval.

\subsubsection{The confidence interval}

For $|y_1| > |y_2|$, the confidence interval for $\theta_1$ is:

\[
   \cI_1(y)= [ \mu_-, \mu_+ ],
\]
where
\[
  \mu_-=\inf\{ a \in \Re: a+c_{(a,0)} \ge y_1 \} 
\]

\[
\mu_+=\sup \{ a \in \Re: a-c_{(a,0)} \le y_1 \} .
\]

Since $|c_{(\mu_{1},0)}| \le c_{(0,0)}$, both endpoints are between
$y_{1}-c_{(0,0)}$ and $y_{1}+c_{(0,0)}$:
the interval is shorter than \v{S}id\'{a}k's simultaneous interval.
For $\alpha = 0.05$,  the interval is widest when $y \approx (\pm2.23, 0)$;
there, the acceptance region for $\mu=0$ just includes $y$. 
The maximum width is about $93.6\%$ of the width of the \v{S}id\'{a}k intervals. 
As $|y_1| \rightarrow \infty$, the length converges to that of the standard unadjusted confidence interval,
about 88\% of the length of the \v{S}id\'{a}k interval.

Unlike the \SoS{} interval for the larger of two, this \SoS{} interval for the larger 
absolute value of two does not automatically work when $\{Y_i\}$ are dependent, because
$\Pr_\mu \{ Y \in B_{\mu, c} \}$ can be lower than it is when they are independent (see \citet{HechtlingerYotam2014CIft}). 
Acceptance regions for parameters with dependent estimators could be calibrated by 
brute force computation,  then inverted computationally to construct confidence intervals.

\section{Largest $k$ of $m$} \label{sec:largest_k_of_m}

Rather than construct a confidence interval for the single parameter estimated to be the larger of $2$,
as in section \ref{sec:max_2},
in this section we construct \SoS{}-controlling confidence intervals corresponding 
for the $k$ parameters estimated to be the largest of $m$ parameters. 

Consider $m$ independent random variables 
$Y = (Y_1, \ldots, Y_m)$.
Let $F_{\theta_i}^i$ be the CDF of $Y_i$, where $F_{\zeta}^i(y) = F_0^i (y-\zeta)$.
Let $Y_{\rank{1}} \ge Y_{\rank{2}} \ge \cdots \ge Y_{\rank{m}}$ be the order statistics
of $\{Y_1, \ldots, Y_m\}$.
Let $y = \left(y_{1},\ldots,y_{m}\right) \in \Re^m$
and let $y_{\rank{1}} \ge y_{\rank{2}} \ge \cdots \ge y_{\rank{m}}$
be the observed order statistics.

Consider the selection rule $S_k(\cdot)$ that keeps the components corresponding to the largest $k$
components of $Y$, with ties broken lexicographically, and let 
$S_k^c(y) \equiv \{1, \ldots, m\} \setminus S_k(y)$.
Let $|s|$ denote the cardinality of the set $s$, so  $|S_k(y)| = k$.
Let $\cS_k \equiv \{ s \subset \{1, \ldots, m\}: |s|=k \}$.

Since $S$ contains the $k$ components for which $Y_i$ is largest, one might expect
the conditional distribution $Y_i$ given $i \in S$ to be
stochastically larger than its unconditional distribution.
It is, which allows us control the chance that the upper endpoint of any interval
is below its parameter using a Bonferroni adjustment with multiplicity $k$ rather than $m$.

Consider the $k$ intervals
\beq
\label{eq:confidence_intervals}
\cI_{\rank{i}}(y) \equiv [y_{\rank{i}}-\ubar{c}_i, y_{\rank{i}}+\bar{c}_i], \;\;1 \le i \le k,
\eeq 
where 
$\ubar{c}_i \equiv (F_0^i)^{-1}(1-\ubar{\lambda})$ and
$\bar{c}_i \equiv -(F_0^i)^{-1}(\bar{\lambda})$, $i = 1, \ldots, m$,
with $\ubar{\lambda} > 0$, $\bar{\lambda} > 0$, and $\ubar{\lambda} + \bar{\lambda} < 1$

\begin{prop}\label{prop:k_out_of_m_coverage}
\beq
    \Pr \left \{ \exists i \in \{1, \ldots, m\}: \cI_{\rank{i}} \not \ni \theta_{\rank{i}} \right \} 
      \le m\ubar{\lambda} + k\bar{\lambda}.
\eeq
\end{prop}

\begin{corollary}\label{cor:family_of_intervals}
For all $\delta \in (0, 1)$, if $\ubar{\lambda} = \delta \alpha/m$ and
$\bar{\lambda} = (1-\delta)\alpha/k$, then 
the intervals defined in equation \ref{eq:confidence_intervals} have $1-\alpha$ \SoS{} coverage. 
\end{corollary}

Corollary \ref{cor:family_of_intervals} defines a family of intervals that control 
\SoS{} at level $\alpha$. 
The intervals are in general asymmetric, and the chance that the intervals miss the parameter
from below is not in general equal to the chance that they miss from above.
For $\delta=\frac{1}{2}$, neither the expected rate at which the upper endpoint is below its parameter
nor the expected rate at which the lower endpoint is above its parameter exceeds
$\alpha/2$.
When $\{ F_0^i \}$ are all symmetric, setting $\delta=m/(m+k)$ gives 
symmetric intervals corresponding to the two sided $m+k$ Bonferroni correction. 

\begin{figure}
\centering
\includegraphics[width=0.7\textwidth]{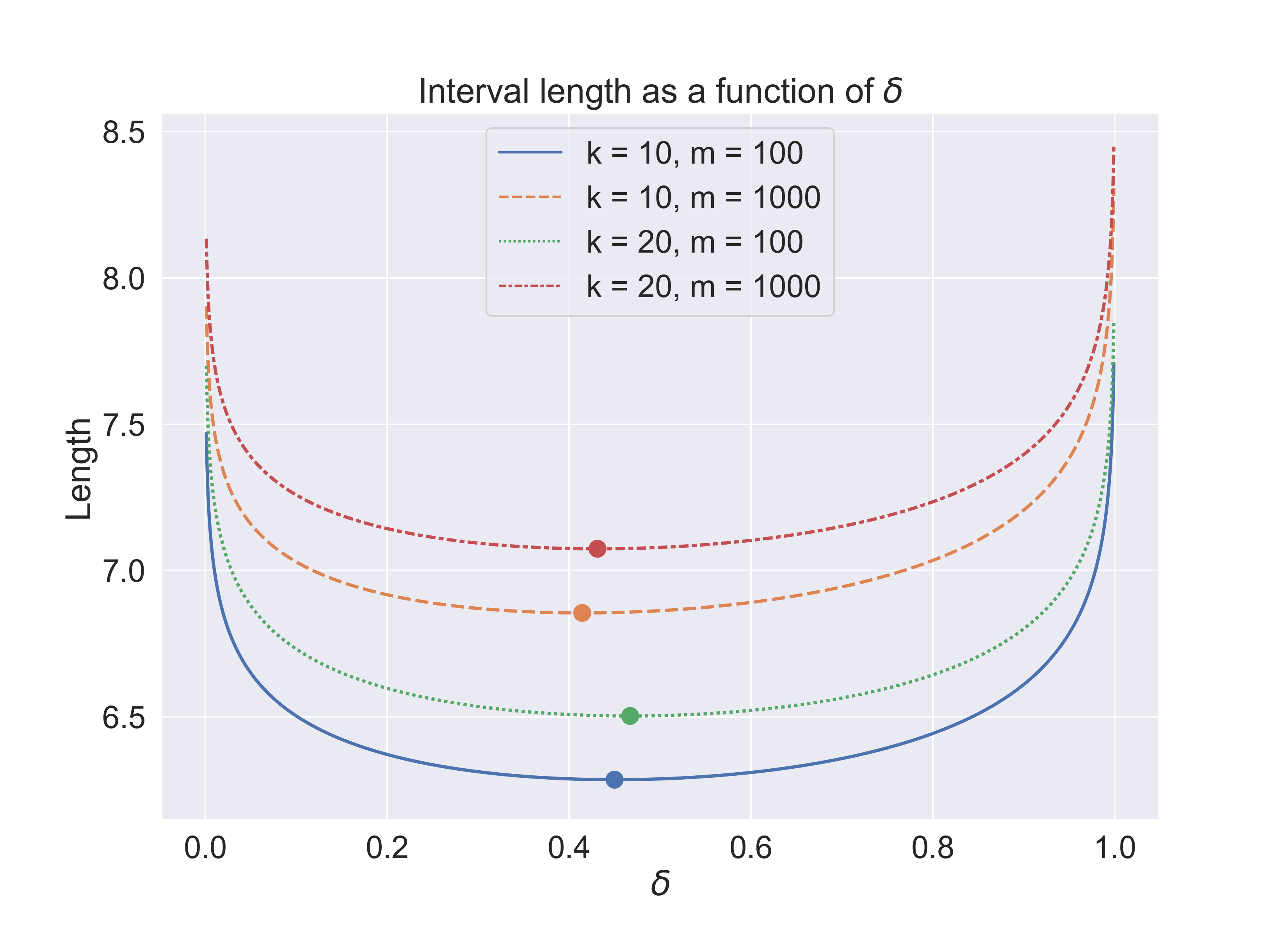}
 
\caption{ \protect \label{fig:delta_impact}
Lengths of the confidence intervals as a function of $k$, $m$, and $\delta$,
for $\alpha=0.05$. 
The dots are plotted at the minimizing values of $\delta$; lengths are nearly minimal when $\delta\approx 0.45$.}
\end{figure}

When $\{ F_0^i \}$ are all equal, finding the $\delta$ that yields the shortest intervals 
is a 1-dimensional optimization problem.
Figure~\ref{fig:delta_impact} plots the length of the intervals for different values of 
$k$ and $m$ when $Y_i \sim N(\theta_i, 1)$. 
Because $S$ selects large components, the shortest intervals extend below $Y_i$ by more than 
they extend above $Y_i$.

\begin{proof}[Proof of Proposition \ref{prop:k_out_of_m_coverage}]
Define
\beq
    \ubar{V}_i=\begin{cases}
    1, & y_i - \ubar{c}_i > \theta_i \\
    0, & \text{otherwise}
    \end{cases};\;\; 
    \bar{V}_i = \begin{cases}
    1, & y_i + \bar{c}_i<  \theta_i\\
    0, & \text{otherwise}
    \end{cases}
\eeq
and
\beq
    \ubar{V}_{\rank{i}}=\begin{cases}
    1, & y_{\rank{i}} - \ubar{c}_i > \theta_{\rank{i}} \\
    0, & \text{otherwise}
    \end{cases};\;\; 
    \bar{V}_{\rank{i}}=\begin{cases}
    1, & y_{\rank{i}} + \bar{c}_i< \theta_{\rank{i}}\\
    0, & \text{otherwise.}
    \end{cases}
\eeq

Now  $\cI_{\rank{i}} \not \ni \theta_{\rank{i}}$ if either 
$Y_{\rank{i}} - \ubar{c}_i > \theta_{\rank{i}}$ (then the lower endpoint of the interval is greater than 
$\theta_{\rank{i}}$, and $\ubar{V}_{\rank{i}} = 1$) or $Y_{\rank{i}} + \bar{c}_i < \theta_{\rank{i}}$ 
(then the upper endpoint of the interval is below $\theta_{\rank{i}}$, and $\bar{V}_{\rank{i}}=1$).

Since   $\ubar{V}_{\rank{i}}=1$ and 
$\bar{V}_{\rank{i}}=1$ are mutually exclusive,
the event $\cI_{\rank{i}} \not \ni \theta_{\rank{i}}$ is the event 
$\ubar{V}_{\rank{i}} + \bar{V}_{\rank{i}}=1$.
Let $\ubar{V} \equiv \sum_{i=1}^k \ubar{V}_{\rank{i}}$ and 
$\bar{V} \equiv \sum_{i=1}^k \bar{V}_{\rank{i}}$.
The event that at least one interval does not cover its parameter is the event
$\ubar{V} + \bar{V} \ge 1$.
Hence,
\beq \label{eq:decompose}
    \Pr \{ \exists i: \cI_{\rank{i}} \not \ni \theta_{\rank{i}} \} = \Pr \left \{ \ubar{V} + \bar{V} \ge 1 \right \} \le 
    \Pr \left \{ \ubar{V} \ge 1 \right \} + \Pr \left \{\bar{V} \ge 1 \right \}. 
\eeq
The first term on the right hand side is
\begin{align}
    \Pr \left\{ \ubar{V} \ge 1 \right \} 
   & \le 
    \sum_{i=1}^k \Pr \left \{ \ubar{V}_{\rank{i}}= 1 \right \} 
    \le 
    \sum_{i=1}^m \Pr \left \{ \ubar{V}_{\rank{i}}= 1 \right \} \nonumber \\
   &  =
    \sum_{i=1}^m \Pr \left \{ \ubar{V}_i  = 1 \right \}   =
    \sum_{i=1}^m \Pr \left \{ Y_i - \theta_i > \ubar{c}_i \right \} \nonumber \\
   & = m \ubar{\lambda}. \label{eq:ubarV}
\end{align}
The second term on the right of \eqref{eq:decompose} is
\begin{align*}
    \Pr \left \{ \bar{V} \ge 1 \right \} 
    & =  \Pr \left \{ \exists i \in S(Y) : \bar{V}_i = 1 \right \} \\
    & = \sum_{s \in \cS_k}  \Pr \left \{ S(Y) = s \cap \exists i \in s : \bar{V}_i = 1 \right \}\\
    & \le \sum_{s \in \cS_k} \sum_{i \in s} \Pr \left \{ S(Y) = s \cap \bar{V}_i = 1 \right \} \\
    & = \sum_{s \in \cS_k} \sum_{i \in s} \Pr \left \{ S(Y) = s \;\mid \; \bar{V}_i = 1 \right \}
          \Pr \{ \bar{V}_i = 1 \} \\
    & = \sum_{s \in \cS_k} \sum_{i \in s} \Pr \left \{ S(Y) = s \; \mid \; \bar{V}_i = 1 \right \}
          \bar{\lambda} \\
    & = \bar{\lambda} \sum_{s \in \cS_k} \sum_{i \in s} \Pr \left \{ \min_{j \in s} Y_j \ge \max_{k \in s^c} Y_k  \; \mid \;  Y_i < \theta_i - \bar{c}_i \right \}
\end{align*}
Let $\tilde{F}_{\theta_i}(y) \equiv \Pr \{ Y_i \le y \mid Y_i < \theta_i - \bar{c}_i \}$.
Then for all $y \in \Re$,
\begin{align*}
   \tilde{F}_{\theta_i}(y)  & =  \frac{\Pr \{ Y_i \le y \cap  Y_i < \theta_i- \bar{c}_i \}}{\Pr \{ Y_i < \theta_i - \bar{c}_i \}} \\
   & = \begin{cases}
       F_{\theta_i}(y)/F_{\theta_i}( \theta_i - \bar{c}_i), & y < \theta_i - \bar{c}_i \\
       1, & y \ge \theta_i - \bar{c}_i 
       \end{cases}
       \\
    & \ge F_{\theta_i}(y).
\end{align*}
By Theorem 4.12.3 of \citet{grimmettStirzaker01}, there exists a 
random variable $\tilde{Y}_i$ independent of $\{Y_j \}_{j \ne i}$ such that
$\tilde{Y}_i \sim \tilde{F}_{\theta_i}$ and 
$$ \Pr \{ \tilde{Y}_i (\omega) \le Y_i(\omega) \}~=~1.$$
Hence,
\begin{align*}
   \Pr \left \{ \min_{j \in s} Y_j \ge \max_{k \in s^c} Y_k  \; \mid \;  Y_i < \theta_i - \bar{c}_i \right \}
   &= \Pr \left \{ \min(Y_i, \min_{j \in s \setminus i} Y_j) \ge \max_{k \in s^c} Y_k  \; \mid \;  Y_i < \theta_i - \bar{c}_i \right \} \\
   &= \Pr \left \{ \min(\tilde{Y}_i, \min_{j \in s \setminus i} Y_j) \ge \max_{k \in s^c} Y_k  \right \} \\
   & \le \Pr \left \{ \min(Y_i, \min_{j \in s \setminus i} Y_j) \ge \max_{k \in s^c} Y_k  \right \} \\
   & = \Pr \left \{ \min_{j \in s} Y_j \ge \max_{k \in s^c} Y_k  \right \} \\
   & = \Pr \{ S(Y) = s \},
\end{align*}
where the first step uses the independence of $\{Y_j\}$.
Thus
\begin{align}
     \Pr \left \{ \bar{V} \ge 1 \right \} 
    & \le \bar{\lambda} \sum_{s \in \cS_k} \sum_{i \in s} \Pr \left \{ S(Y) = s \right \}\nonumber \\
    & = \bar{\lambda} \sum_{s \in \cS_k} k  \Pr \left \{ S(Y) = s \right \}\nonumber \\
    & = k \bar{\lambda}. \label{eq:barV}
\end{align}
Combining \eqref{eq:decompose}, \eqref{eq:ubarV}, and \eqref{eq:barV} yields the desired result.
\end{proof}
\section{Comparisons} \label{sec:Comparison}

Bonferroni confidence intervals are of the form $[Y_j - c_, Y_j + c]$, with
\beq
   c =  -\sup \{ c: \Pr \{ Y_i-\theta_i \le c \} \le \alpha/(2m) \}.
\eeq
Bonferroni confidence intervals control \SoP{} (and consequently \SoS{}) even when
$\{Y_i\}$ are dependent. 
When $\{Y_i\}$ are independent, \v{S}id\'{a}k confidence intervals,
which are of the same form but with
\beq
   c =  -\sup \{ c: \Pr \{ Y_i-\theta_i \le c \} \le (1 - (1-\alpha)^{1/m})/2\},
\eeq
offer a small (but optimal) improvement on Bonferroni intervals.

\citet{fuentes2018confidence} construct \SoS{} intervals for this $k$ out of $m$ problem
with $\{Y_i-\theta_i\}$ are IID $N(0,1)$.
Their intervals, which we call \emph{FCW}, are of the form 
$[Y_j - c, Y_j + d]$, where $c$ and $d$ are constants such that

\beq
	\left[\Phi\left(c\right)-\Phi\left(-d\right)\right]^{k-1} \times \left[\Phi^{m-k+1}\left(c\right)-\Phi^{m-k+1}\left(-d\right)\right] \ge 1-\alpha.
\eeq

\begin{figure}[ht]
\centering
\includegraphics[width=0.85\textwidth]{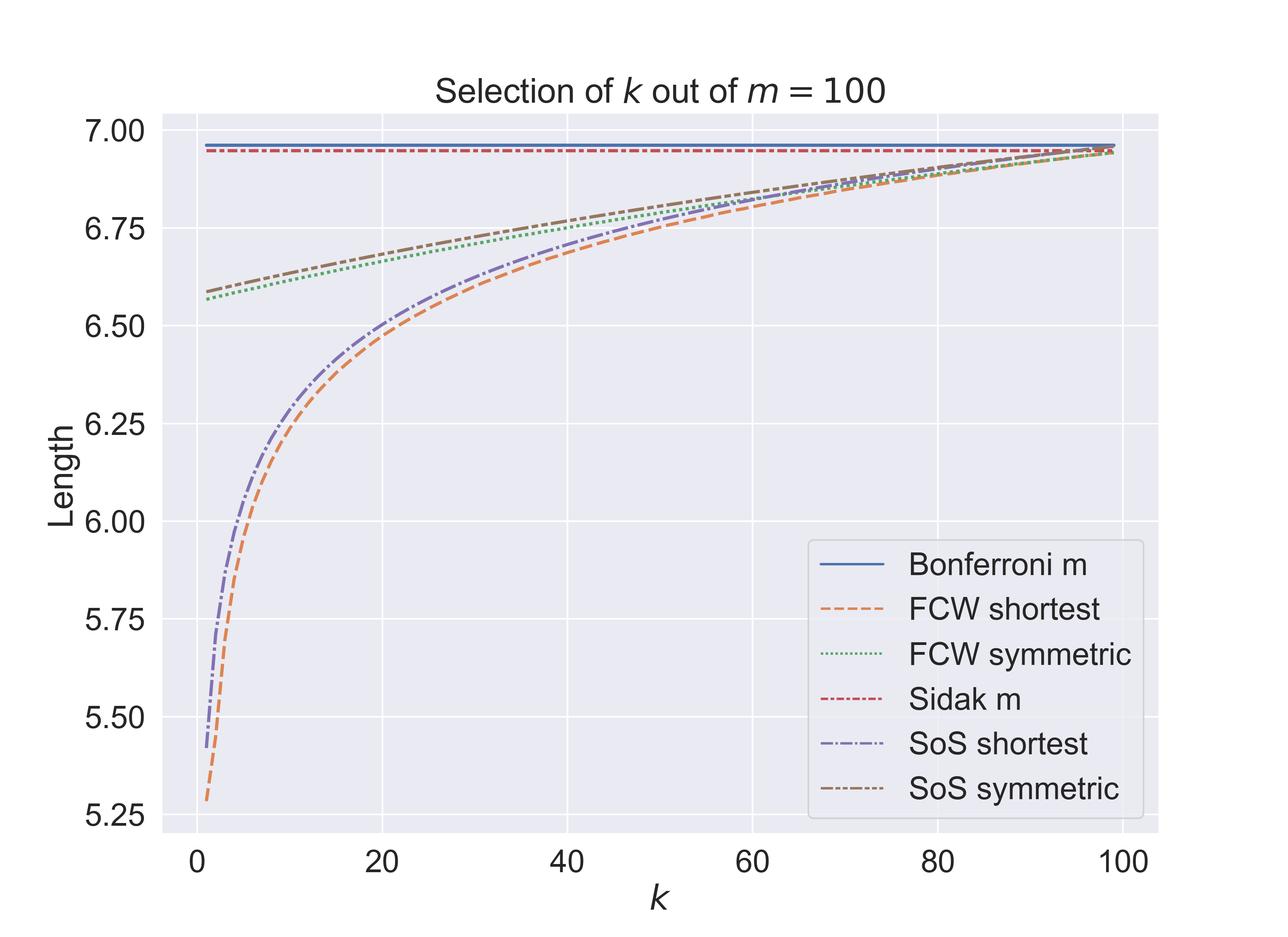}
\caption{ \protect \label{fig:intervals_comparisons}
Lengths of confidence intervals with simultaneous coverage of the $k$ of $m=100$ 
normal means estimated to be largest, for $\alpha=0.05$.}
\end{figure}

Figure~\ref{fig:intervals_comparisons} compares 
the Bonferroni intervals, the \v{S}id\'{a}k intervals,
FCW intervals (with $c=d$ and with $c$ and $d$ chosen to make the intervals as
short as possible),
and the \SoS{} intervals (with $\delta$ chosen to make the intervals symmetric or as short as possible).

Because \SoS{} intervals make inferences about only $k<m$ parameters, one would
expect them to be narrower than \SoP{} intervals---but not as narrow as intervals for a
pre-specified set $k$ parameters: there should be some penalty for
the selection.
(The case $m=2$, $k=1$ is a remarkable exception.)
Letting the intervals be asymmetric can reduce their length further.
Suppose all the parameters are equal. 
Because $S$ selects the largest $k$, $S$ will
preferentially select components for which $Y_i > \theta_i$.
Using intervals that extend below $Y_i$ more
than they extend above $Y_i$ allows the intervals to be shorter.
Figure~\ref{fig:intervals_comparisons} shows that the advantage of allowing asymmetry is 
substantial when $k \ll m$.
The \SoS{} intervals are slightly wider than the FCW intervals, but the FCW intervals 
require $\{Y_i - \theta_i\}$ to have normal distributions, while the \SoS{} intervals do not.

\citet{weinstein2013selection}, 
\citet{fithian2014optimal} and \citet{reid2017post}
develop \CoS{} intervals for the largest $k$ of $m$.
Controlling \CoS{} controls \FCR{}. 
These intervals also control \SoS{} if each interval is constructed at level $1-\alpha/|S|$,
rather than $1-\alpha$.

\CoS{} intervals use conditional distributions, which
depend on the conditioning event and the underlying parameters
of the model. 
As a result, \CoS{} intervals are sometimes useful
and sometimes not. 
For instance, \CoS{} intervals for the largest of
$m$ or largest absolute value of $m$ can be too wide to be useful when
the largest and second-largest estimators are close \citep{fithian2015selective}. 
Table~\ref{table:CoS_comparison} shows the \textit{uniformly most powerful unbiased}
(UMPU) \CoS{} intervals from example~4 in \citet{fithian2014optimal} when $\left(Y_{1},Y_{2}\right)$
is bivariate uncorrelated normal centered at $\left(\theta_{1},\theta_{2}\right)$, 
for the observed value $\left(y_{1},y_{2}\right)=\left(2.9,2.5\right)$
and the greater $S\left(Y\right)=\left\{ \left(1\right)\right\}$ is selected.
In section~\ref{sec:larger_of_2}, we showed that the unadjusted univariate interval has the right
coverage. 
The UMPU saturated \CoS{} intervals use $\Pr_{\theta_{1}}\left(Y_{1}|Y_{1}>2.5\right)$.
For $\theta_{1}\ll2.5$, the conditional distribution is a renormalized tail of the distribution,
which yields long intervals. 
The UMPU selected
\CoS{} intervals use $\Pr_{\left(\theta_{1},\theta_{2}\right)}\left(Y_{1}\mid Y_{1}>Y_{2}\right)$.
Then, the underlying distribution is a function of both $\theta_{1}$
and $\theta_{2}$. 
Setting $\theta_{2}=0$ produces ``nicer'' intervals than $\theta_{2}=2.5$
because the selection event has less effect on the conditional
distribution. 
But while setting the coefficient of an unselected variable
to zero in a model might make sense, for location parameters, setting
$\theta_{2}=0$ when it is estimated to be $2.5$ does not. 
This shows the importance of the unselected parameters to the coverage of selected parameters.
Furthermore, if a coverage were needed for all $\theta_{2}$, the interval
would have an infinite lower tail, since for all $\theta_{1}\ll2.5$ there exist $\theta_{2}\gg2.5$ such that $\Pr_{\left(\theta_{1},\theta_{2}\right)}\left(Y_{1}|Y_{1}>Y_{2}\right)$
is centered at $2.9$. 

Because the lengths of \CoS{} intervals depend on the observed value of $Y$ (not only on $k$ and $m$), we do not include \CoS{} intervals in our figures or numerical comparisons.

\begin{table}
\centering
\begin{tabular}{|l|c|c|}
\hline 
 & $\theta_{1}\in\left(\text{Lower},\text{Upper}\right)$ & Length\tabularnewline
\hline 
\hline 
Marginal / SoS & $\left(0.940,4.859\right)$ & $3.919$\tabularnewline
\hline 
$\Pr_{\theta_{1}}\left(Y_{1}|Y_{1}>2.5\right)$ - UMPU saturated & $\left(-9.090,4.594\right)$ & $13.684$\tabularnewline
\hline 
$\Pr_{\left(\theta_{1},0\right)}\left(Y_{1}|Y_{1}>Y_{2}\right)$ - UMPU
selected $\left(\theta_{2}=0\right)$ & $\left(0.772,4.855\right)$ & $4.082$\tabularnewline
\hline 
$\Pr_{\left(\theta_{1},2.5\right)}\left(Y_{1}|Y_{1}>Y_{2}\right)$ -
UMPU selected $\left(\theta_{2}=2.5\right)$ & $\left(-0.224,4.558\right)$ & $4.782$\tabularnewline
\hline 
$\Pr_{\left(\theta_{1},\theta_{2}\right)}\left(Y_{1}|Y_{1}>Y_{2}\right)$
- UMPU selected $\left(\forall\theta_{2}\in\Re\right)$ & $\left(-\infty,4.859\right)$ & $\infty$\tabularnewline
\hline 
\end{tabular}
\caption{Intervals for $\alpha=0.05$ for the bivariate normal after selecting the greater of the two
estimators when $\left(y_{1},y_{2}\right)=\left(2.9,2.5\right)$. 
The depicted UMPU saturated and selected \CoS{} intervals are from \citet{fithian2014optimal}.}
\label{table:CoS_comparison}
\end{table}

\section{Discussion} 

\citet{katsevich2018towards} address ``simultaneous selective inference'' in testing.
They consider making selective inference statements on many selection rules  
$\left\{ S_{i}\left(Y\right)\right\}$, guaranteeing this statements hold simultaneously 
with high probability.
By restricting the possible selections to those determined by a particular algorithm,
for instance, those obtained by varying the level at which FDR is controlled, their
method can improve on full simultaneity.

We have shown that knowing the selection rule $S$ can improve inferences,
for instance by allowing one to control \SoS{} or \CoS{} without necessarily controlling \SoP{}.
More information is better.
\citet{lee2016exact} and \citet{tibshirani2016exact} 
constructed confidence interval for parameters 
selected by the Lasso and by forward stepwise selection, respectively. 

For \SoP{} and \FCR{}, it is not necessary to know $S$, but if $S$ is known,
improvements are possible.
\citet{berk2013valid} addressed the problem of inference when the model is selected
because a pre-specified explanatory variable had the highest statistical significance, 
which restricts the family over which simultaneous coverage is required. 
The restriction transforms the problem into that of making a confidence interval for the coefficient
whose estimate is most significantly nonzero among many correlated estimates.
This amounts to selecting on the basis of the largest 1 of $m$ correlated statistics,
which can be solved computationally for problems of modest dimension.
If the design matrix is orthogonal, their confidence interval amounts to
the Bonferroni interval adjusted for $m$ parameters---the dimension of the relevant family.
\citet{weinstein2013selection} constructed \CoS{} intervals tailored to avoid containing zero,
at the expense of being somewhat longer where it does not matter and \citet{weinstein2014selective} designed \FCR{} intervals that try to avoid covering 0.  
\citet{barber2015controlling} took advantage of the fact that their knockoff method
used forward stepwise search (their method does not yet offer confidence intervals).

To see that \FCR{} intervals can also take advantage of knowledge of $S$, consider
the rule $S$ that selects the $k$ parameters estimated to be the largest of $m$.
This rule always selects $k$ parameters. 
The proof of proposition~\ref{prop:k_out_of_m_coverage} bounds the probability of making one 
or more errors via the expected number of upper endpoints that are smaller than their corresponding
parameter plus the expected number of lower endpoints that are larger than their corresponding
parameters. 
These expectations are bounded by a multiple of $\alpha$. 
\FCR{} control requires the expected number of non-covering intervals, divided by $k$, to be at most $\alpha$.
Therefore, if $S$ selects the $k$ largest of $m$, \FCR{} intervals can have lower endpoints
based on the $1-(\alpha / 2)\cdot(k/m)$ quantile (which works for any \emph{simple} selection rule
\citep{benjaminiYekutieli05}),
and upper endpoints based on the  $\alpha/2$ quantile instead of the $(\alpha / 2)\cdot (k/m)$ quantile.

In conclusion, specification of a selection rule allows us to explore the inference ground between simultaneous  and false confidence statement rate controlling confidence intervals, by defining the in-between goal of simultaneous coverage one the selected. 
We explored the implications of this error-rate by studying the Òlargest k-out-of-mÓ rule. 
This resulted in new confidence intervals. In their simplest version, the lower endpoints are Bonferroni adjusted for m, 
while the upper endpoints are Bonferroni adjusted only for k, offering uniform improvement over regular Bonferroni.
Furthermore, the utilization of such a selection rule allows improvement over the general FCR intervals.

It seems clear that the confidence intervals retain their coverage for positive-regression-dependent 
estimators, because the  conditional distribution of the variables selected is stochastically greater
than the same variables without selection, but we do not have a formal proof. 
Appendix~\ref{app:dependency_simulations} gives numerical evidence that this is indeed the case. 
The intervals may also be valid when $k$ is random, for instance when selection
depends on crossing a fixed threshold or some other testing procedure.

We think the results for the $k$ largest in absolute value of $m$ extend to $m>2$ and $k>1$, 
but again, we offer no formal proof. 
We have decided to publish our current results without those generalizations with the hope that they will
spark interest engage others in pursuing this approach.
\appendix
\section{Proof of Proposition \ref{prop:max_abs_of_2_bound}}\label{app:max_abs_of_2_proof}

Since $Y_i - \theta_i$ are IID standard normals, they are rotationally invariant and symmetric.
Of course, $\Pr_\mu \{A\} = \Pr_0 \{ A - \mu\}$.
These three properties let us find or bound $\Pr \{B_{\mu,c} \}$.

Assume without loss of generality that  $mu_1 \ge mu_2 \ge 0$.
The acceptance regions $B_{\mu, c}$ have three forms, depending on whether
(1)~$\mu_1\ge \mu_2\ge  c$, 
(2)~$\mu_1\ge  c\ge \mu_2$, 
or
(3)~$c\ge \mu_1\ge \mu_2$.
\\[1ex]

\begin{figure}
\centering
\begin{subfigure}{.38\textwidth}
  \centering
  \includegraphics[width=1\textwidth]{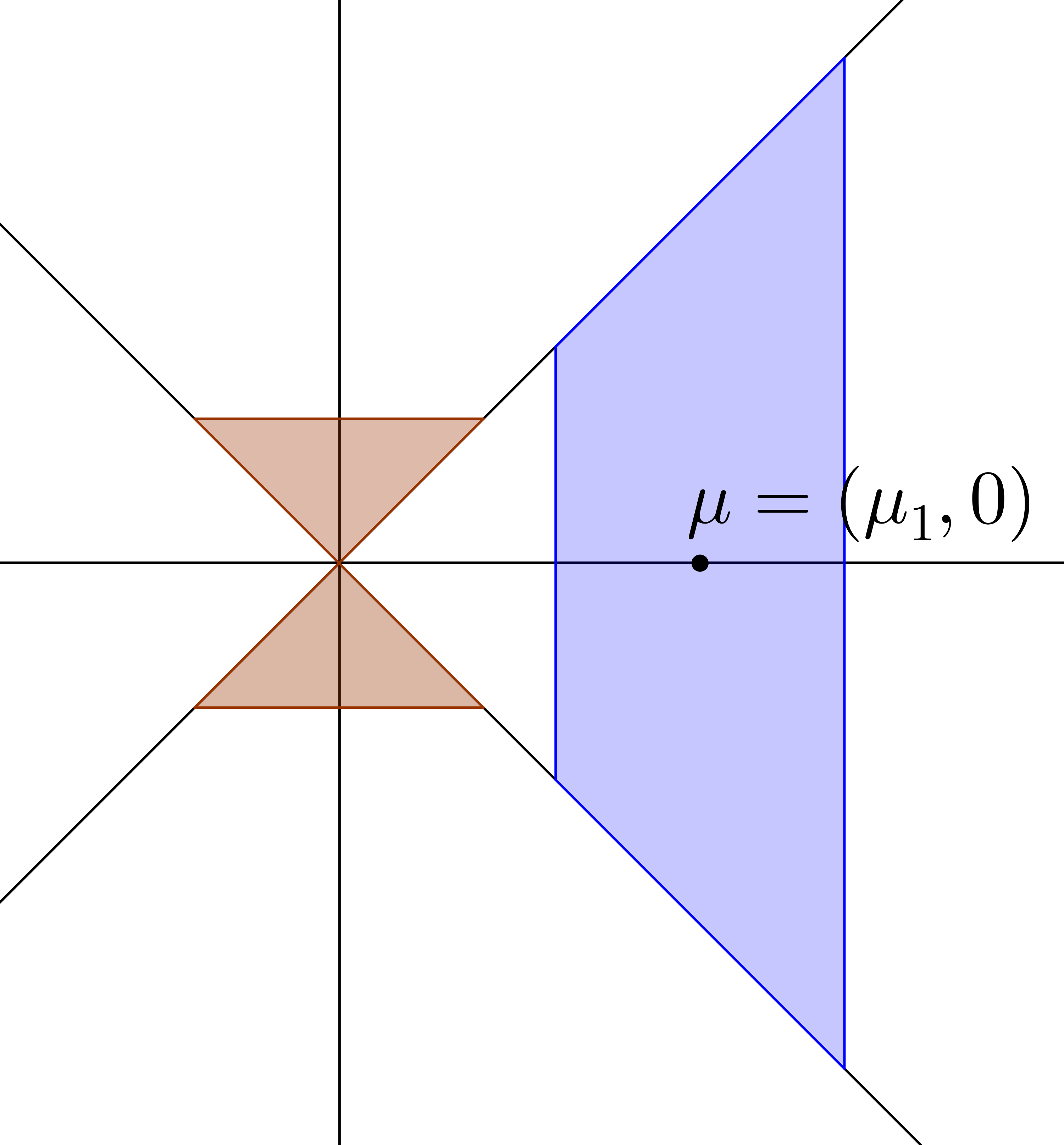}
  \caption{}
\end{subfigure}%
\begin{subfigure}{.38\textwidth}
  \centering
  \includegraphics[width=1\textwidth]{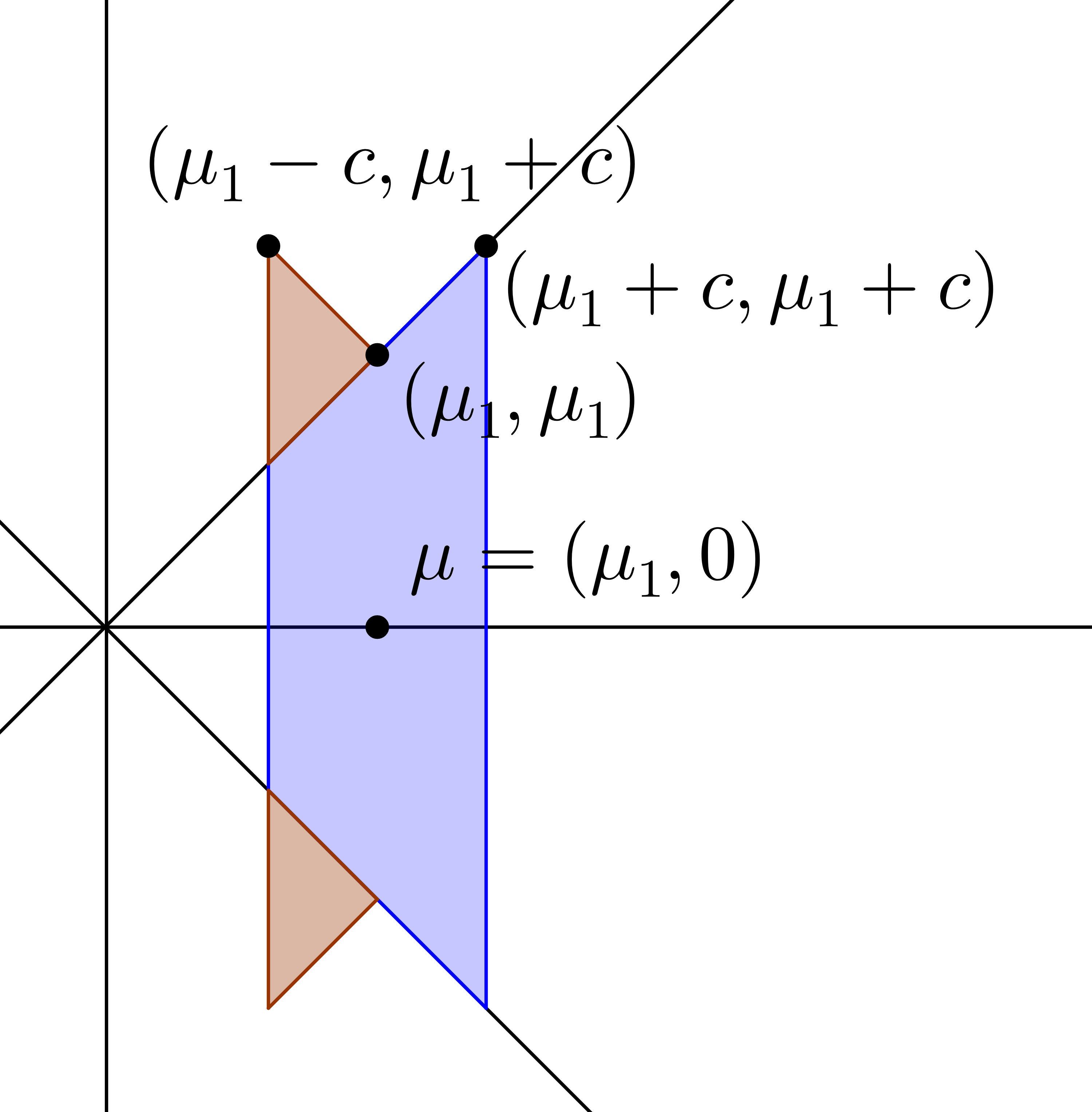}
  \caption{}
\end{subfigure}
\caption{ \protect \label{fig:first_B} (a): $B_{(\mu_1,0),c}$. (b): $B_{(\mu_1,0),c}$
after a $\Pr_\mu$-preserving transformation.}
\end{figure}

\begin{figure}
\centering
\begin{subfigure}{.42\textwidth}
  \centering
  \includegraphics[width=1\textwidth]{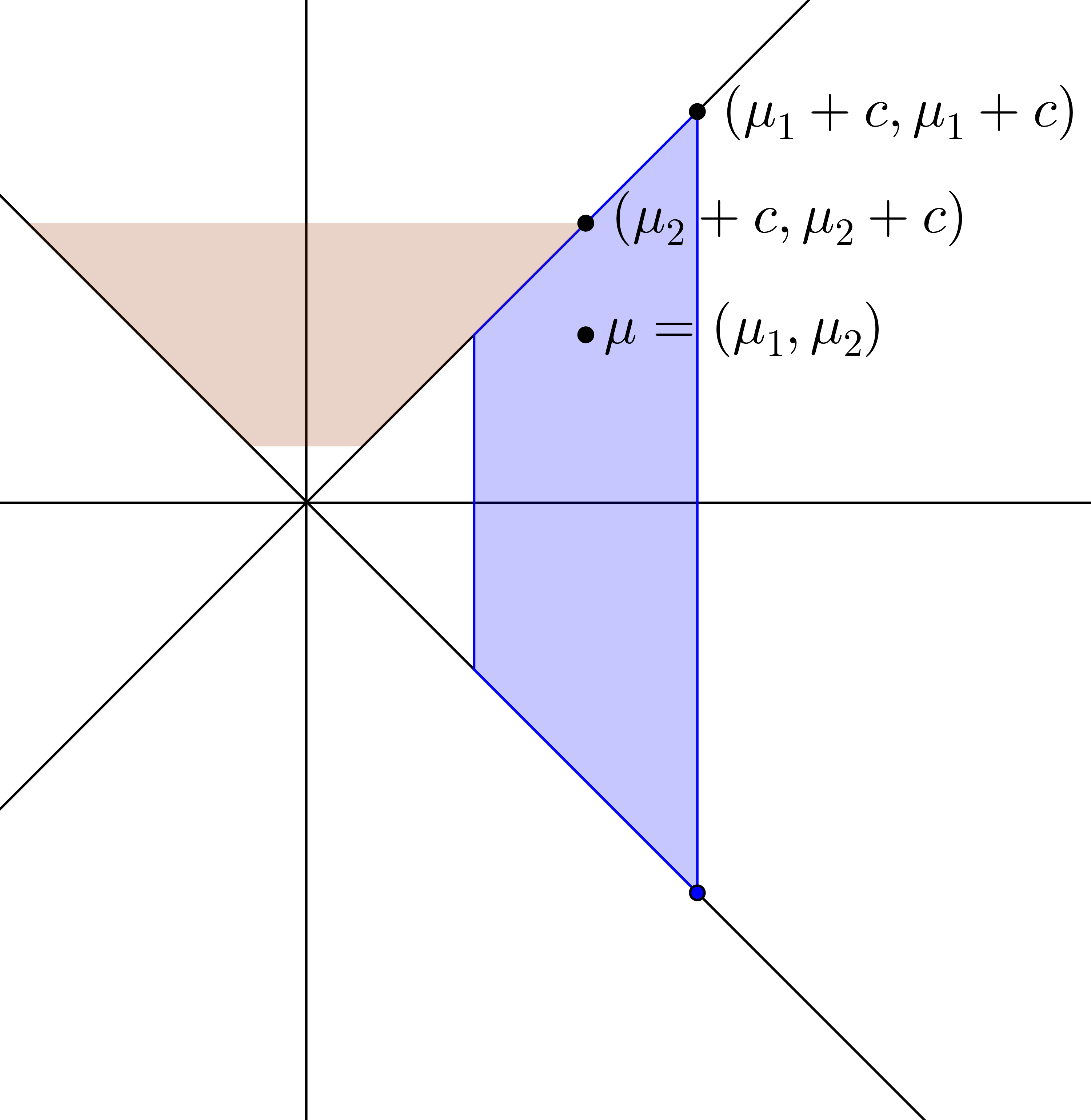}
  \caption{}
\end{subfigure}%
\begin{subfigure}{.42\textwidth}
  \centering
 \includegraphics[width=1\textwidth]{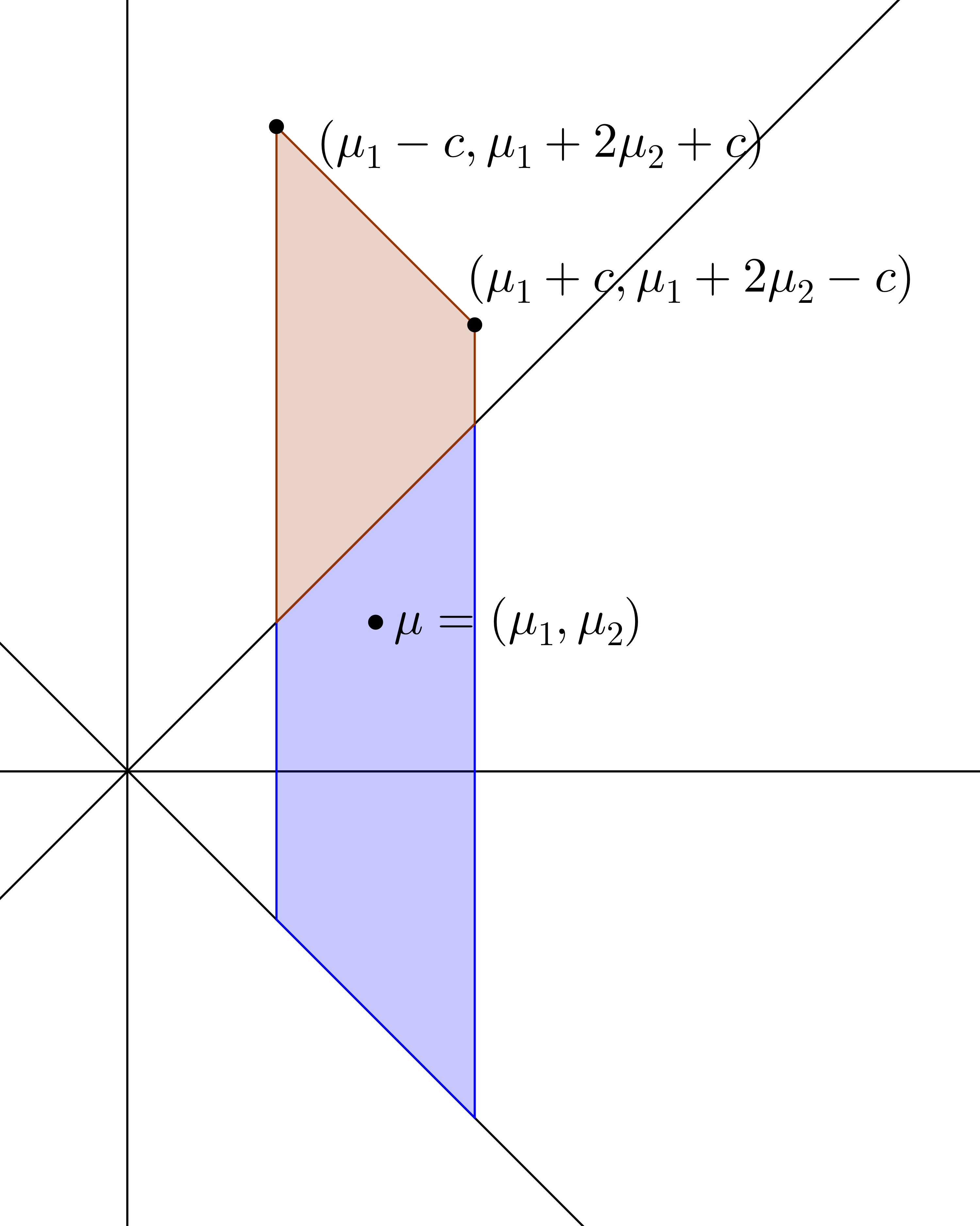}
  \caption{}
\end{subfigure}
\caption{\protect \label{fig:second_B} (a):~$B_{(\mu_1,\mu_2),c}$ for $\mu_1 > \mu_2 > c$. 
(b):~$B_{(\mu_1,\mu_2),c}$ after a
$\Pr_\mu$-preserving transformation.}
\end{figure}

\begin{figure}
\centering
\begin{subfigure}{.42\textwidth}
  \centering
  \includegraphics[width=1\textwidth]{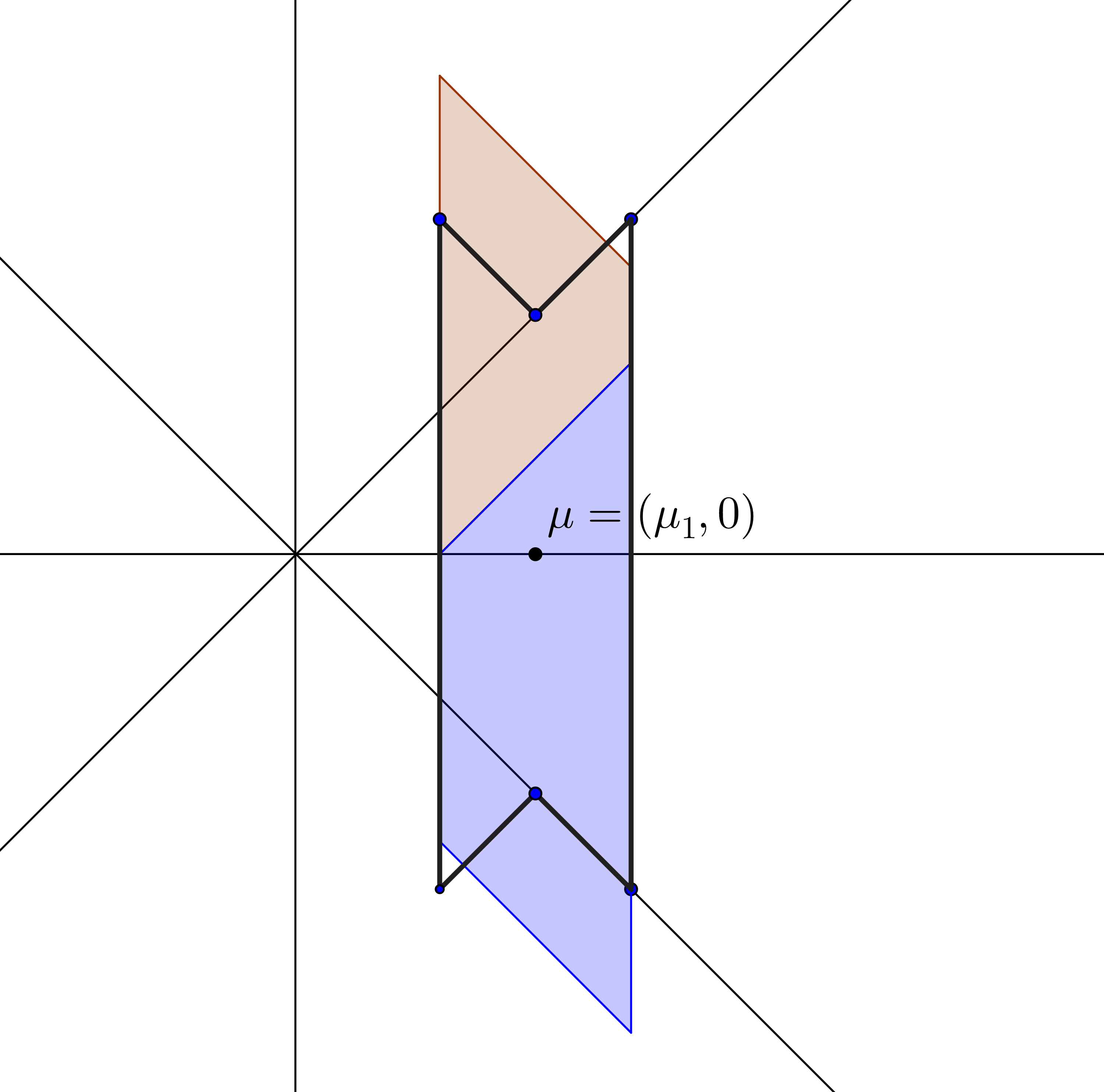}
  \caption{ }
\end{subfigure}%
\begin{subfigure}{.42\textwidth}
  \centering
  \includegraphics[width=1\textwidth]{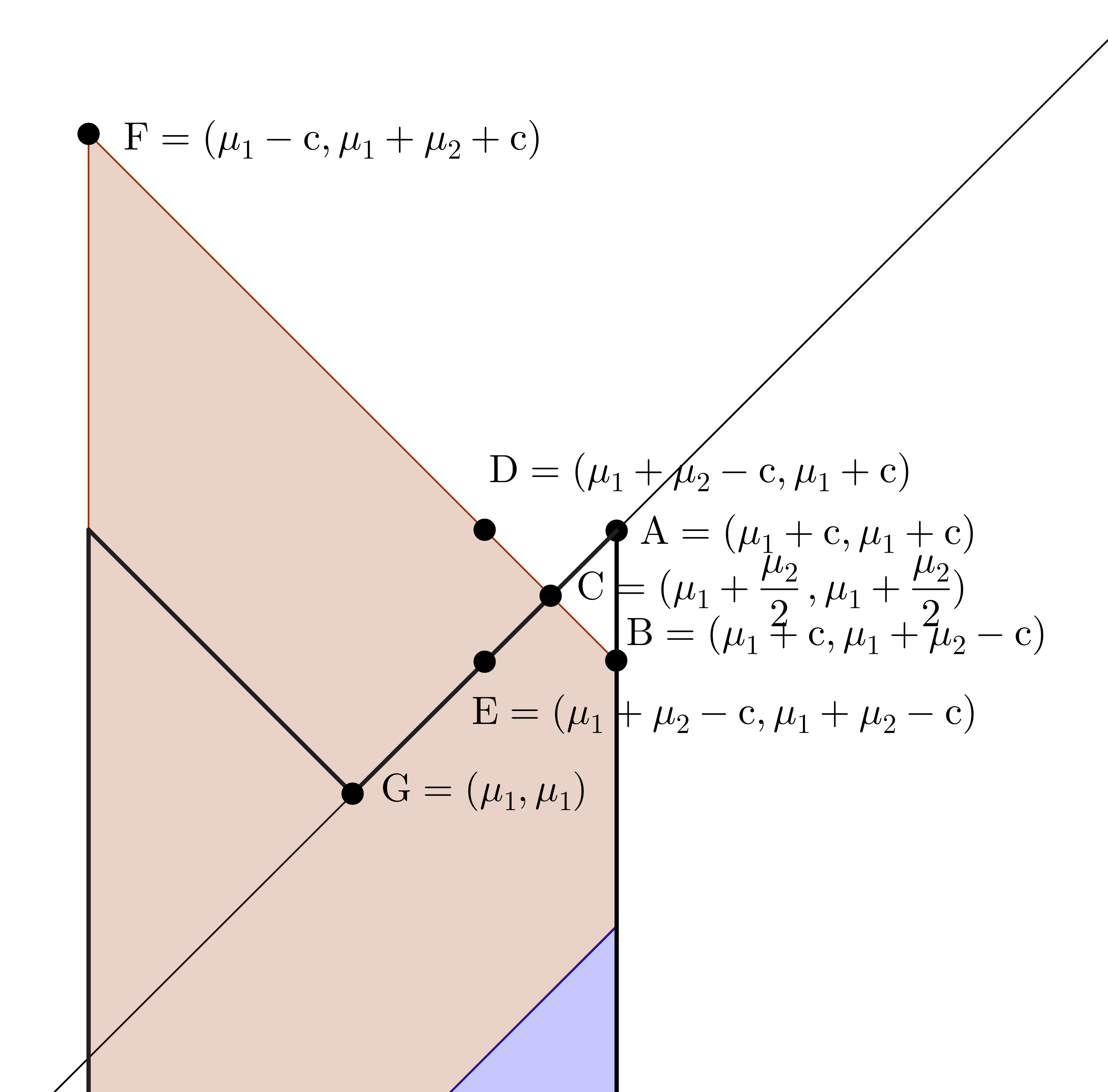}
  \caption{ }
\end{subfigure}
\caption{ \protect \label{fig:third_B} 
(a):~$B_{(\mu_1,\mu_2),c}$ and $B_{(\mu_1,0),c}$ superposed
after $\Pr_\mu$-preserving transformations when $\mu_1\ge \mu_2\ge  c$.
The black hexagon is $B_{(\mu_1,0),c}$
and the shaded parallelogram is $B_{(\mu_1,\mu_2),c}$ after
re-centered at $(\mu_1,0)$. 
(b):~Detail of the top of panel~(a). 
The bottom half of the trapezoid is the top half rotated by $\pi$, which does
change its probability.}
\end{figure}

\noindent
\textbf{Case (1)} $\mu_1\ge \mu_2\ge  c$.
Because $\mu_1\ge  c$,  $B_{(\mu_1,0),c}^1$ is a trapezoid
and $B_{(\mu_1,0),c}^2$ consists of two congruent triangles. 
Rotating the triangles and reflecting them about $x=\mu_1$
yields the hexagon in Figure~\ref{fig:first_B}.
Because $\mu_2\ge  c$, $B_{(\mu_1,\mu_2),c}$ consists of two trapezoids. 
Clockwise rotation and reflection of the trapezoid $B_{(\mu_1,\mu_2),c}^2$
yields a parallelogram. See Figure~\ref{fig:second_B}.
The transformation $\varphi(x,y)=(x,y-\mu_2)$ translates $(\mu_1,\mu_2)$
to $(\mu_1,0)$, allowing us to compare their acceptance
regions. See Figure~\ref{fig:third_B}. 

The line $\overline{FB}$ in Figure~\ref{fig:third_B} (b) is $y=-x+2\mu_1+\mu_2$.
It intersects $y=x$ at the point $C=\left(\mu_1+ \mu_2/2, \mu_1+ \mu_2/2\right)$.
If $\mu_2/2 \ge  c$, then $B_{(\mu_1,0),c} \subseteq B_{(\mu_1,\mu_2),c}$. Otherwise, $D=(\mu_1+\mu_2-c,\mu_1+c)$
is above the point $E=(\mu_1+\mu_2-c,\mu_1+\mu_2-c)$, and
to the right of $G=(\mu_1,\mu_1)$. 
Thus the triangles
$\triangle CDE$ and $\triangle ABC$ are congruent. 
Since the reflection of $\triangle CDE$ about the horizontal line $x=\mu_1+\mu_2/2$
is $\triangle ABC$, $\Pr \{ \triangle CDE\} \ge  \Pr \{\triangle ABC\}$,
and so $\Pr_{(\mu_1,\mu_2)} \{ A_{(\mu_1,\mu_2), c} \} \ge  
\Pr_{(\mu_1,0)} \{A{}_{(\mu_1,0),c} \}$.
\\[1ex]

\begin{figure}
\centering
\begin{subfigure}{.41\textwidth}
  \centering
  \includegraphics[width=1\textwidth]{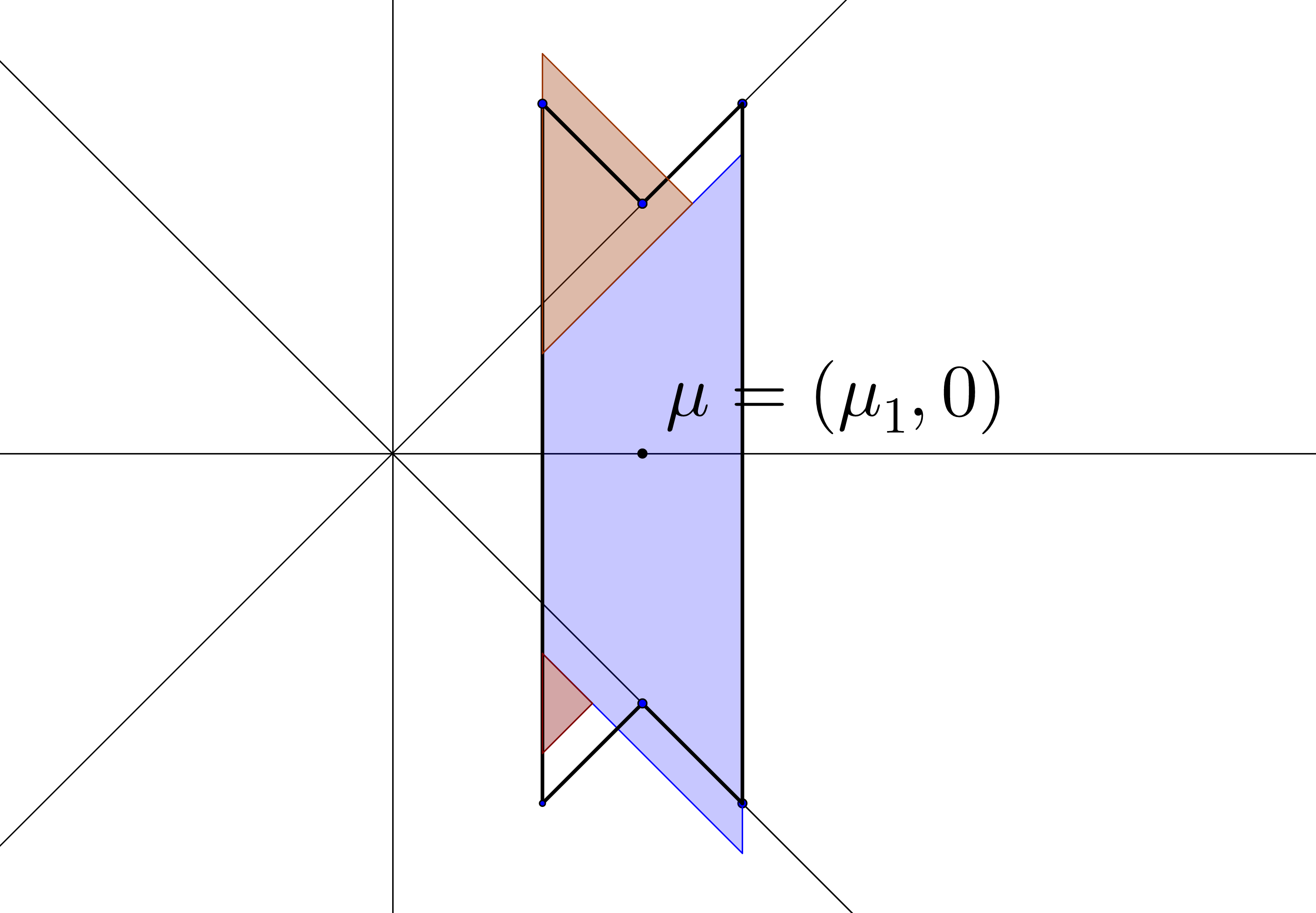}
  \caption{ }
\end{subfigure}%
\begin{subfigure}{.41\textwidth}
  \centering
  \includegraphics[width=1\textwidth]{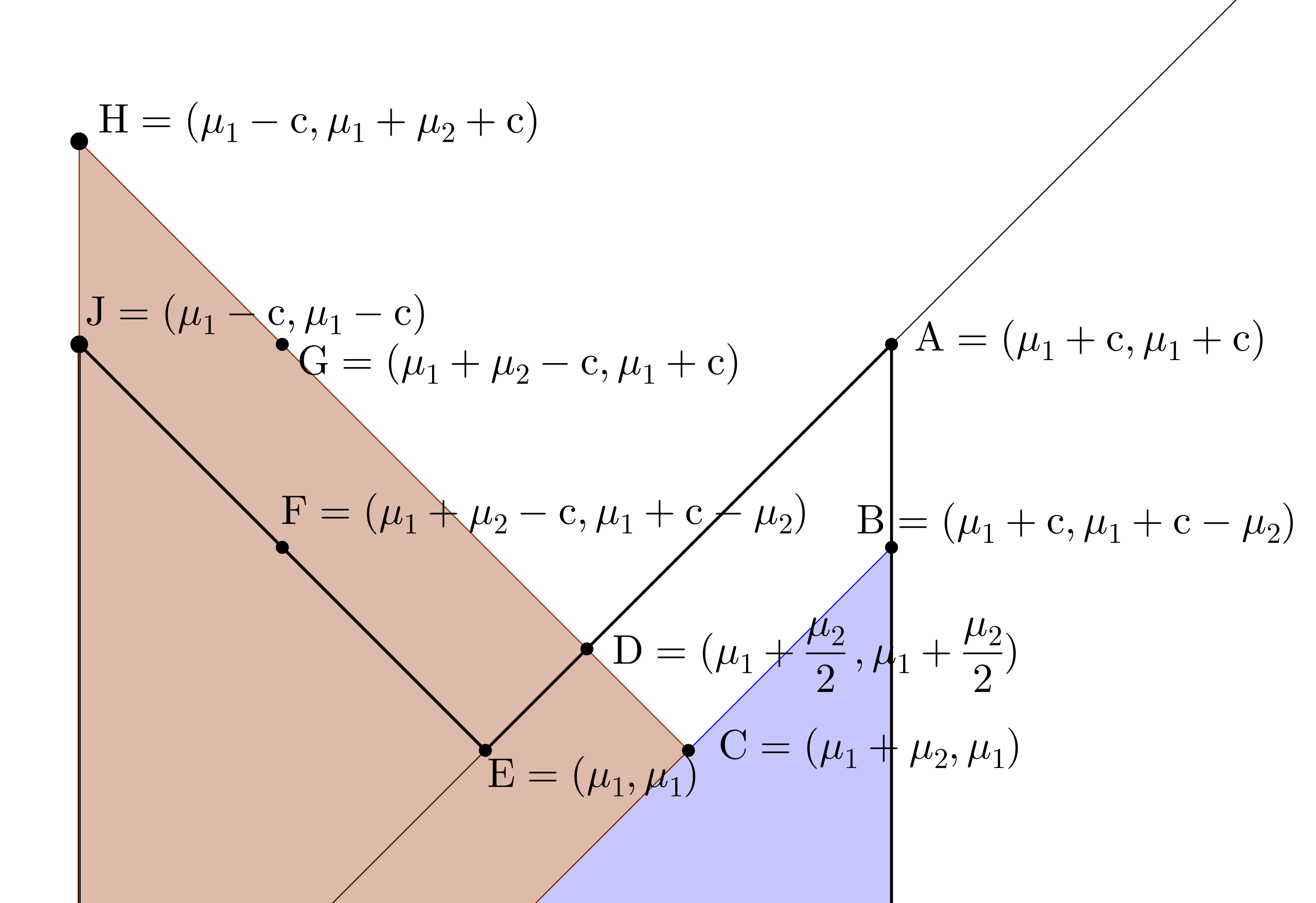}
  \caption{ }
\end{subfigure}
\caption{ \protect \label{fig:fourth_B} 
(a):~$B_{(\mu_1,\mu_2),c}$ for $\mu_1 > c > \mu_2$ superposed with
$B_{(\mu_1,0),c}$, after a $\Pr_\mu$-preserving transformations.
(b):~Detail of the top of panel~(a).}
\end{figure}

\noindent
\textbf{Case (2)} $\mu_1\ge  c\ge \mu_2$.
The region $B_{(\mu_1,0),c}$ is as in case~(1),
but for $\mu_2 > 0$, $B_{(\mu_1,\mu_2),c}^1$ is a trapezoid
and $B_{(\mu_1,\mu_2),c}^2$ consists of two triangles that meet
at 0.
Rotating the upper triangle $\pi/2$ clockwise, and the lower
triangle $\frac{\pi}{2}$ counter clockwise, then reflecting
both across $x=\mu_1$ and centering yields Figure~\ref{fig:fourth_B}.
As in case \textbf{(1)}, rotating the
quadrilateral $DEFG$ about the line $x=\mu_1+\mu_2/2$,
and noticing that once again the top half is congruent to the bottom
half (rotated by $\pi$) establishes the conclusion. 
\\[1ex]

\begin{figure}
\centering
\begin{subfigure}{.41\textwidth}
  \centering
  \includegraphics[width=1\textwidth]{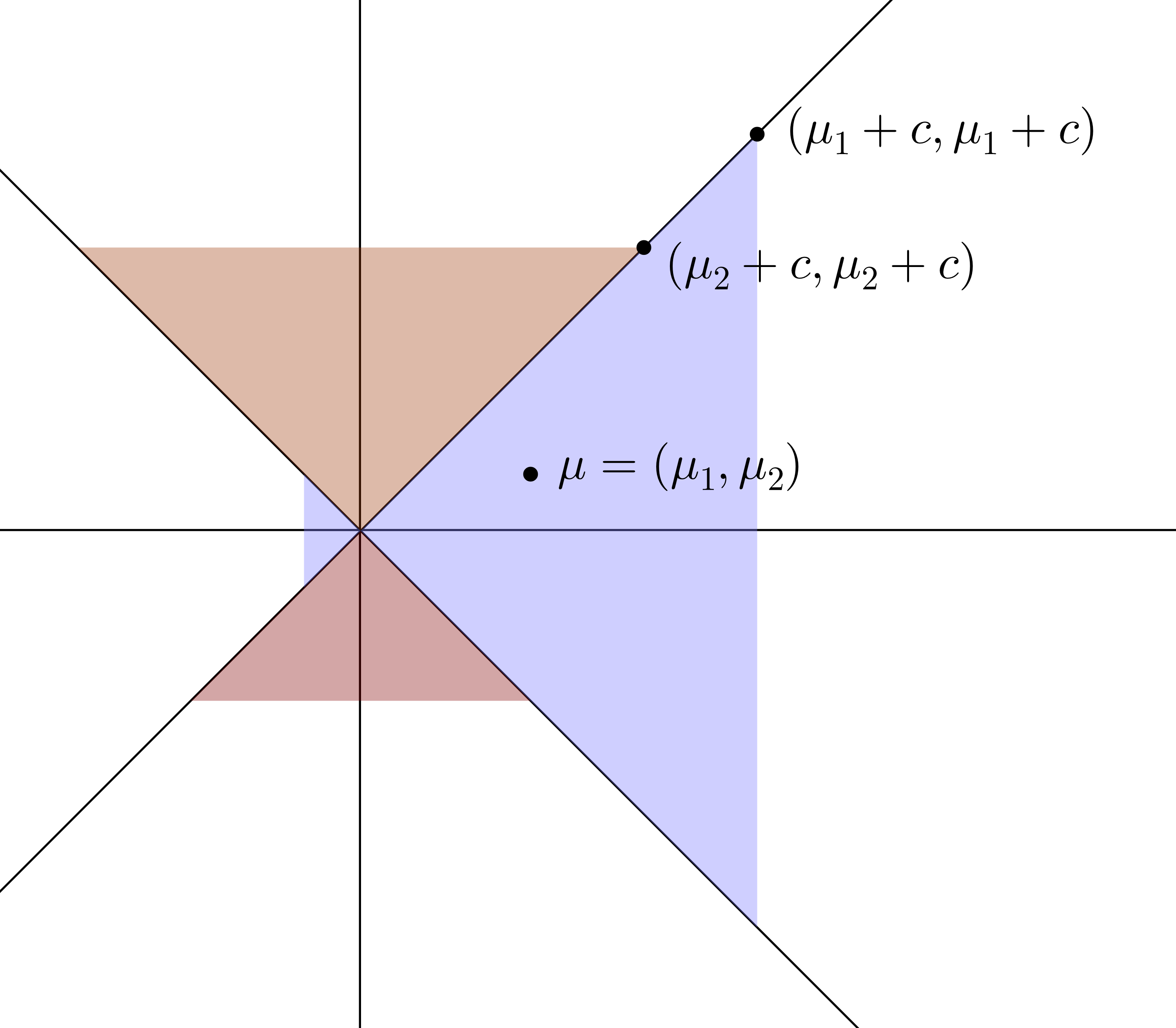}
  \caption{ }
\end{subfigure}
\begin{subfigure}{.41\textwidth}
  \centering
  \includegraphics[width=1\textwidth]{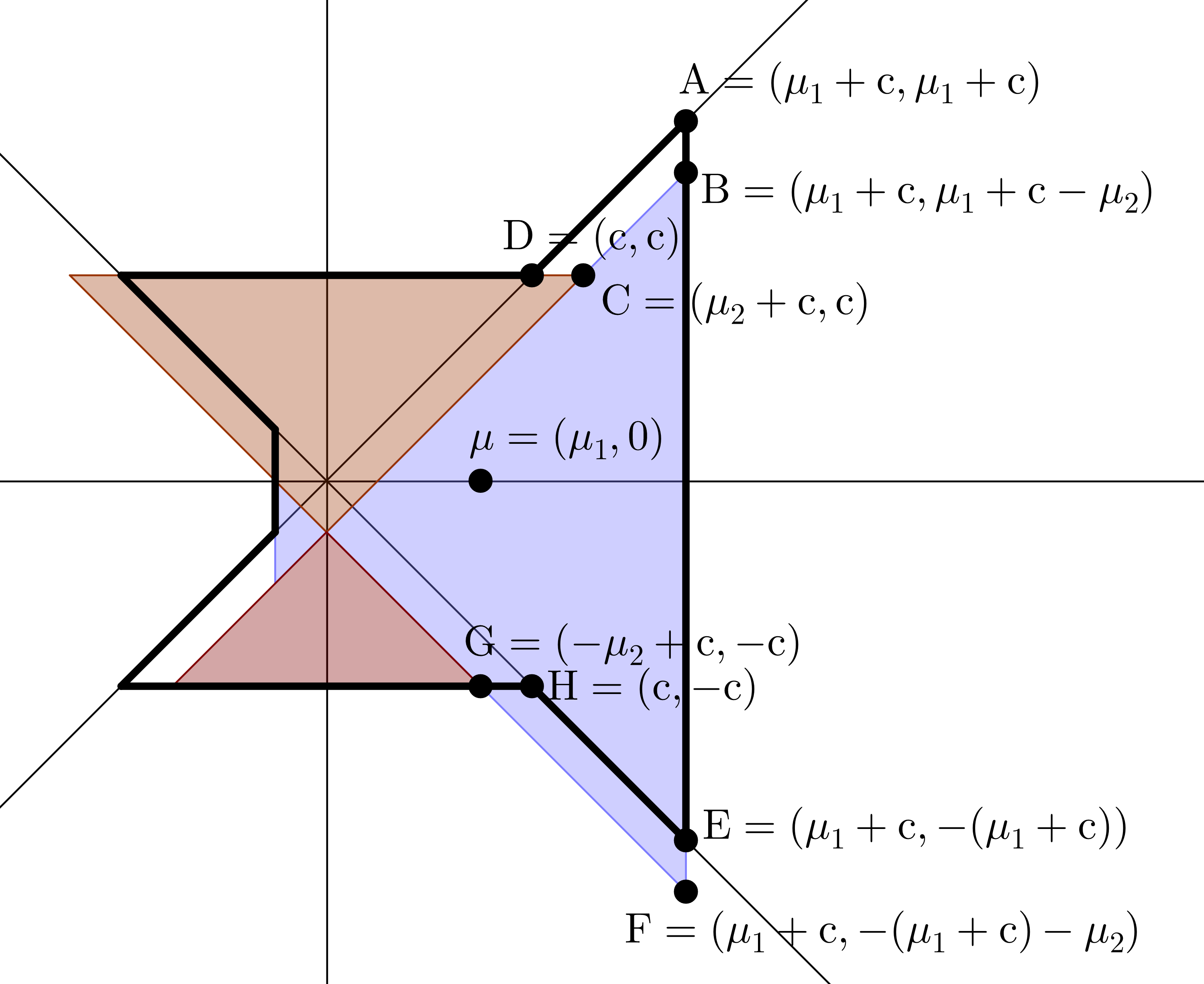}
  \caption{ }
\end{subfigure}
\caption{\protect \label{fig:fifth_B} $B_{(\mu_1,\mu_2),c}$ and $B_{(\mu_1,0),c}$
when $c \ge \mu_1 \ge \mu_2$. (a):~$B_{(\mu_1,\mu_2),c}$. (b):~$B_{(\mu_1,0),c}$ 
(the black polygon) and 
 $B_{(\mu_1,\mu_2),c}$
(the shaded polygon) after the translation $h(x)=x-\mu_2$. }
\end{figure}

\noindent 
\textbf{Case (3)} $c\ge \mu_1\ge \mu_2$.
The only transformation required to compare the probabilities of the regions 
is to translate $B_{(\mu_1,\mu_2),c}$ to 
center it at $\left(\mu_1,0\right)$.
See Figure~\ref{fig:fifth_B}.
The area
of the quadrilateral $EFGH$ in Figure~\ref{fig:fifth_B}(b) is greater than
the area of the quadrilateral $ABCD$, and $EFGH$
is closer in norm to $\left(\mu_1,0\right)$, so
$\Pr_{(\mu_1,\mu_2)}\{A_{(\mu_1,\mu_2),c}\}\ge  \Pr_{(\mu_1,0)}\{A_{(\mu_1,0),c}\}$. 

\section{Dependency Simulations}\label{app:dependency_simulations}
The conditional distribution of the variables selected is stochastically
greater than the same variables independently of the selection. This
gives rise to the intuition that the method should work under positive
dependency. We test the intuition in simulations. Following \cite{cai2014two} 
we simulate each time $50,000$ observations and consider $m=100$, $k=10$. To demonstrate the methods is applicable when the variables have different distributions, we sample $50$ variables of $Y$ (denoted as $Y_{\left[1:50\right]}$) from multivariate normal $Y_{\left[1:50\right]}\sim\mathcal{N}\left(\theta_{\left[1:50\right]}\cdot\eta,\Sigma\right)$, and $50$ variables (denoted as $Y_{\left[51:100\right]}$) from multivariate $t$ with $5$ degrees of freedom, $Y_{\left[51:100\right]}\sim t_{5}\left(\theta_{\left[51:100\right]}\cdot\eta,\Sigma\right)$. In each simulation we use the same $\Sigma$ for both the normals and the $t$ variables. For all variables, $\theta_{i}\sim\text{Uniform}\left(-1,1\right)$ and $\eta$ transition
in the different simulations from $0$ (identically distributed) to
$40$ (large differences). We consider three
types of $\Sigma$ covariance structures.

\begin{enumerate}
\item Auto Regression (AR) matrix such that $\Sigma=\left(\sigma_{i,j}\right)$,
and each entry $\sigma_{i,j}=\rho^{\left|i-j\right|}$ for $1\leq i,j\leq m$
and $\rho\in\left\{ 0.3,0.7\right\} .$
\item Time Decay model (TD) which taken from model $7$ in \cite{cai2014two}. $\Sigma^{\star}=\left(\sigma_{i,j}^{\star}\right)$,
$\sigma_{ij}^{\star}=\left|i-j\right|^{-5}/2$ for $j\not=i$. The
covariance matrix is $\Sigma=D^{1/2}\Sigma^{\star}D^{1/2}$, where
$D=\left(d_{i,j}\right)$, such that $d_{i,j}\sim\text{Uniform}\left(1,3\right).$ 
\item Block Covariance matrix (Block) - $5$ blocks sized $10$ each, with
$\rho\in\left\{ 0,0.2,0.5,0.75,0.9\right\} .$
\end{enumerate}

\begin{figure}
\centering
\includegraphics[width=1\textwidth]{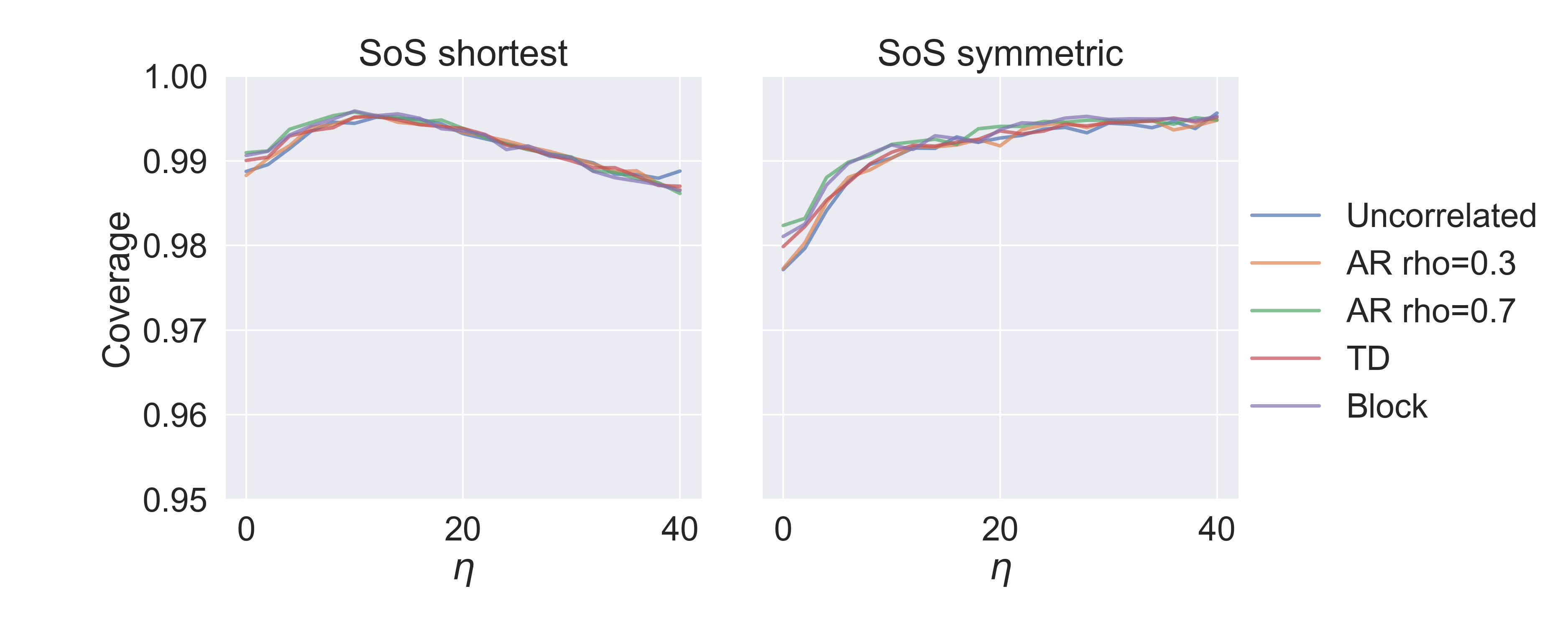}
\caption{ \protect \label{fig:intervals_dependency_simulations}
Intervals simultaneous coverage for different dependency structures. The selection of $k=10$ of $m=100$ estimated to be the largest out of $50$ normal and $50$ $t_5$ correlated variables for $\alpha=0.05$.}
\end{figure}

Figure \ref{fig:intervals_dependency_simulations} shows the coverage of the SoS shortest and symmetrical intervals
with different structures of dependency for $\alpha=0.05$. All intervals
obtain the desired coverage and often too conservative, demonstrating
the bound might be improved in particular cases. In addition, affect
of the different $\Sigma$ is relatively minor and always more conservative.
The major affect on the coverage is the selection distribution since the results vary significantly as a function of $\eta$.

\eject
\bibliographystyle{abbrvnat}
\bibliography{pbsBib}

\begin{thebibliography}{23}
\providecommand{\natexlab}[1]{#1}
\providecommand{\url}[1]{\texttt{#1}}
\expandafter\ifx\csname urlstyle\endcsname\relax
  \providecommand{\doi}[1]{doi: #1}\else
  \providecommand{\doi}{doi: \begingroup \urlstyle{rm}\Url}\fi

\bibitem[Barber and Cand{\`e}s(2015)]{barber2015controlling}
R.~F. Barber and E.~J. Cand{\`e}s.
\newblock Controlling the false discovery rate via knockoffs.
\newblock \emph{The Annals of Statistics}, 43\penalty0 (5):\penalty0
  2055--2085, 2015.

\bibitem[Benjamini and Yekutieli(2005)]{benjaminiYekutieli05}
Y.~Benjamini and Y.~Yekutieli.
\newblock False discovery rate controlling confidence intervals for selected
  parameters.
\newblock \emph{J. Am. Stat. Assoc.}, 100\penalty0 (469):\penalty0 71--80,
  2005.

\bibitem[Berk et~al.(2013)Berk, Brown, Buja, Zhang, and Zhao]{berk2013valid}
R.~Berk, L.~Brown, A.~Buja, K.~Zhang, and L.~Zhao.
\newblock Valid post-selection inference.
\newblock \emph{The Annals of Statistics}, 41\penalty0 (2):\penalty0 802--837,
  2013.

\bibitem[Braun(1994)]{braun1994collected}
H.~I. Braun.
\newblock \emph{The collected works of John W. Tukey: Multiple comparisons,
  1948-1983}.
\newblock Chapman \& Hall, 1994.

\bibitem[Cai et~al.(2014)Cai, Liu, and Xia]{cai2014two}
T.~T. Cai, W.~Liu, and Y.~Xia.
\newblock Two-sample test of high dimensional means under dependence.
\newblock \emph{Journal of the Royal Statistical Society: Series B (Statistical
  Methodology)}, 76\penalty0 (2):\penalty0 349--372, 2014.

\bibitem[Fithian et~al.(2014)Fithian, Sun, and Taylor]{fithian2014optimal}
W.~Fithian, D.~Sun, and J.~Taylor.
\newblock Optimal inference after model selection.
\newblock \emph{arXiv preprint arXiv:1410.2597}, 2014.

\bibitem[Fithian et~al.(2015)Fithian, Taylor, Tibshirani, and
  Tibshirani]{fithian2015selective}
W.~Fithian, J.~Taylor, R.~Tibshirani, and R.~Tibshirani.
\newblock Selective sequential model selection.
\newblock \emph{arXiv preprint arXiv:1512.02565}, 2015.

\bibitem[Fuentes et~al.(2018)Fuentes, Casella, and
  Wells]{fuentes2018confidence}
C.~Fuentes, G.~Casella, and M.~T. Wells.
\newblock Confidence intervals for the means of the selected populations.
\newblock \emph{Electronic Journal of Statistics}, 12\penalty0 (1):\penalty0
  58--79, 2018.

\bibitem[Grimmett and Stirzaker(2001)]{grimmettStirzaker01}
G.~R. Grimmett and D.~R. Stirzaker.
\newblock \emph{Probability and Random Processes}.
\newblock {Oxford University Press}, August 2001.
\newblock ISBN 0198572220.

\bibitem[Hechtlinger(2014)]{HechtlingerYotam2014CIft}
Y.~Hechtlinger.
\newblock Confidence interval for the selected.
\newblock Master's thesis, Tel Aviv University, Tel Aviv, Israel, June 2014.
\newblock Available at
  \url{https://tau-primo.hosted.exlibrisgroup.com/primo-explore/fulldisplay?docid=972TAU_ALMA71237141730004146&context=L&vid=TAU&search_scope=TAU_Blended&tab=default_tab&lang=en_US},.

\bibitem[Hochberg and Tamhane(1987)]{hochberg1987multiple}
Y.~Hochberg and A.~C. Tamhane.
\newblock \emph{Multiple comparison procedures}.
\newblock Wiley, New York., 1987.

\bibitem[Hsu(1981)]{hsu81}
J.~Hsu.
\newblock Simultaneous confidence intervals for all distances from the best.
\newblock \emph{Ann. Stat.}, 9:\penalty0 1026Ð1034, 1981.

\bibitem[Hsu(1996)]{hsu96}
J.~Hsu.
\newblock \emph{Multiple Comparisons: Theory and Methods}.
\newblock Chapman and Hall, London, 1996.

\bibitem[Katsevich and Ramdas(2018)]{katsevich2018towards}
E.~Katsevich and A.~Ramdas.
\newblock Towards" simultaneous selective inference": post-hoc bounds on the
  false discovery proportion.
\newblock \emph{arXiv preprint arXiv:1803.06790}, 2018.

\bibitem[Lee et~al.(2016)Lee, Sun, Sun, and Taylor]{lee2016exact}
J.~D. Lee, D.~L. Sun, Y.~Sun, and J.~E. Taylor.
\newblock Exact post-selection inference, with application to the lasso.
\newblock \emph{The Annals of Statistics}, 44\penalty0 (3):\penalty0 907--927,
  2016.

\bibitem[Reid et~al.(2017)Reid, Taylor, and Tibshirani]{reid2017post}
S.~Reid, J.~Taylor, and R.~Tibshirani.
\newblock Post-selection point and interval estimation of signal sizes in
  gaussian samples.
\newblock \emph{Canadian Journal of Statistics}, 45\penalty0 (2):\penalty0
  128--148, 2017.

\bibitem[Taylor and Tibshirani(2015)]{taylor2015statistical}
J.~Taylor and R.~Tibshirani.
\newblock Statistical learning and selective inference.
\newblock \emph{Proceedings of the National Academy of Sciences}, 112\penalty0
  (25):\penalty0 7629--7634, 2015.

\bibitem[Tibshirani et~al.(2016)Tibshirani, Taylor, Lockhart, and
  Tibshirani]{tibshirani2016exact}
R.~J. Tibshirani, J.~Taylor, R.~Lockhart, and R.~Tibshirani.
\newblock Exact post-selection inference for sequential regression procedures.
\newblock \emph{Journal of the American Statistical Association}, 111\penalty0
  (514):\penalty0 600--620, 2016.

\bibitem[Tukey(1953)]{tukey1953theProblem}
J.~W. Tukey.
\newblock \emph{The problem of multiple comparisons.}
\newblock Unpublished manuscript. In The Collected Works of John W. Tukey VIII.
  Multiple Comparisons: 1948–1983, 1–300. Chapman and Hall, New York.,
  1953.

\bibitem[Venter(1988)]{venter1988confidence}
J.~Venter.
\newblock Confidence bounds based on the largest treatment mean.
\newblock \emph{South African Journal of Science}, 84\penalty0 (5):\penalty0
  340, 1988.

\bibitem[Wasserstein and Lazar(2016)]{wasserstein2016ASA}
R.~L. Wasserstein and N.~A. Lazar.
\newblock The asa's statement on p-values: Context, process, and purpose.
\newblock \emph{The American Statistician}, 70\penalty0 (2):\penalty0 129--133,
  2016.

\bibitem[Weinstein and Yekutieli(2014)]{weinstein2014selective}
A.~Weinstein and D.~Yekutieli.
\newblock Selective sign-determining multiple confidence intervals with fcr
  control.
\newblock \emph{arXiv preprint arXiv:1404.7403}, 2014.

\bibitem[Weinstein et~al.(2013)Weinstein, Fithian, and
  Benjamini]{weinstein2013selection}
A.~Weinstein, W.~Fithian, and Y.~Benjamini.
\newblock Selection adjusted confidence intervals with more power to determine
  the sign.
\newblock \emph{Journal of the American Statistical Association}, 108\penalty0
  (501):\penalty0 165--176, 2013.

\end{thebibliography}

\end{document}